\documentclass[aps,prx,twocolumn,superscriptaddress,longbibliography]{revtex4-2}
\usepackage[pages=all, color=black, position={current page.south}, placement=bottom, scale=1, opacity=1, vshift=5mm]{background}
\SetBgContents{
	\tt 
}      

\usepackage{amsmath}
\usepackage{amsthm}
\usepackage{amssymb}
\usepackage{times}
\usepackage{braket}
\usepackage{ulem}

\usepackage{tikz}
\usetikzlibrary{quantikz2} 

\usepackage[utf8]{inputenc}
\usepackage{hyperref}

\usepackage{physics}

\usepackage{graphicx, color}
\graphicspath{{fig/}}

\usepackage{algorithm, algpseudocode} 
\usepackage{mathrsfs} 
\usepackage{lipsum}

\begin{document}

\title{Extracting the photon indistinguishability error from measurable quantum observables}

\author{Franciscus~H.~B.~Somhorst}
\email{f.h.b.somhorst@utwente.nl}
\affiliation{MESA+ Institute for Nanotechnology, University of Twente, P.~O.~box 217, 7500 AE Enschede, The Netherlands} 
\author{Jason~Saied}
\affiliation{QuAIL, NASA Ames Research Center, Moffett Field, CA 94035, USA}
\author{Eleanor~G.~Rieffel}
\affiliation{QuAIL, NASA Ames Research Center, Moffett Field, CA 94035, USA}
\author{Jelmer~J.~Renema}
\affiliation{MESA+ Institute for Nanotechnology, University of Twente, P.~O.~box 217, 7500 AE Enschede, The Netherlands}

\begin{abstract}
    We present a method to extract the photon indistinguishability error from Hong-Ou-Mandel interference measurements, accounting for the combined effects of loss and multiphoton noise that contaminate the single-photon Hilbert space. Our analysis resolves apparent inconsistencies in previous interpretations of such measurements. The reported method applies to a wide range of single-photon sources, including quantum dots.
\end{abstract}

\maketitle

\section{Introduction}
\label{sec:intro}
Single-photon states are key resources for linear optics-based quantum information processing \cite{slussarenko2019photonic,flamini2018photonic,couteau2023applications}. Consequently, many underlying system architectures require the ability to generate numerous copies of pure and indistinguishable photons. However, realistic single-photon sources (SPSs) are subject to imperfections, often creating noisy copies that can significantly impact the overall system performance \cite{rohde2006error, renema2018efficient,saied2024advancing,shaw2023errors}.

Photon imperfections are reflected in a range of experimentally measurable quantum observables. The photon-number impurity is usually quantified by the measured second-order intensity autocorrelation at zero time delay, $g^{(2)}(0)$ \cite{glauber1963quantum, grunwald2019effective}, while the degree of photon indistinguishability is usually quantified by the measured Hong-Ou-Mandel (HOM) interference visibility, $V_{\text{HOM}}$ \cite{hong1987measurement,marshall2022distillation}. Unfortunately, these two quantities are not completely independent: even when the photons are fully indistinguishable, the presence of multiphoton noise reduces the perceived visibility. Across the extensive literature on SPSs, the visibility correction factors used to account for photon-number impurity errors vary considerably and often lack a clear physical justification \cite{trivedi2020generation,somaschi2016near,wang2019towards,grange2017reducing,kirvsanske2017indistinguishable,tomm2021bright,ollivier2021hong,gonzalez2025two,ding2025high}.  

In this work, we derive a measurement-dependent visibility correction factor for SPSs in which photon-number impurity primarily results from the leakage of a weak, separable noise field. Our approach is based on a model of an imperfect single-photon state that is broadly applicable to a wide class of SPSs based on single quantum emitters, including quantum dots \cite{michler2000quantum}, Rydberg ensembles \cite{ornelas2020demand}, two-dimensional materials \cite{fournier2023two}, and nitrogen-vacancy centers \cite{bernien2013heralded}. Furthermore, our findings confirm that the correction factor is strongly dependent on the degree of indistinguishability between signal and noise photons \cite{ollivier2021hong,gonzalez2025two}. 
We verify that this dependence is not primarily driven by interference involving three or more photons under typical high-loss experimental conditions. Instead, it arises predominantly from the combined effects of loss and photon-number impurity errors, which transfer a fraction of the multiphoton states into the single-photon subspace. We find that this process leads to an \textit{effective} indistinguishability error that may exceed the intrinsic indistinguishability error \cite{marshall2022distillation}, highlighting the nontrivial interaction between various photon imperfections. 

The paper is structured as follows. We begin by summarizing the key findings of this work and then proceed to substantiate them in detail. First, we introduce a general model of an imperfect photon state that captures key experimental imperfections: loss, partial distinguishability, and multiphoton contributions. 
Next, we derive the resulting expression for the single-photon trace purity $\Tr[\rho_1^2]$, which quantifies how identical the single photon states are. Here, we introduce the concept of the \textit{effective} indistinguishability error, which extends the conventional definition by accounting for interactions between loss and photon-number impurity errors. Finally, we return to our central question: how do we extract the indistinguishability error from measurable quantum observables $V_{\text{HOM}}$ and $g^{(2)}(0)$?

\section{Summary of key findings}
The first the key finding of this work is that the interplay between loss and multiphoton errors can be described as an \textit{effective} indistinguishability error $\tilde{\epsilon}$, such that $\Tr[\rho_1^2] = (1 - \tilde{\epsilon})^2$ defines the single-photon trace purity. For a wide range of practical sources, the effective indistinguishability error is, to first order in $g^{(2)}(0)$, approximated by:
\begin{equation}
    \tilde{\epsilon} = \epsilon + \frac{1}{2}(1-\eta)(1-\epsilon)g^{(2)}(0),
\end{equation} where $\epsilon$ is the intrinsic indistinguishability error and $\eta$ is the overall transmission efficiency.

The second key finding is that the choice of measurement method determines how the measured visibility should be interpreted to extract the photon indistinguishability error:
\begin{itemize}
    \item Use $V_{\text{HOM}} = (1-\tilde{\epsilon})^2 \left(1- g^{(2)}(0) \right)$ when the visibility is extracted from \textit{coincidence counts} \cite{tomm2021bright,gonzalez2025two}. 
    \item Use $V_{\text{HOM}} = (1-\tilde{\epsilon})^2 - g^{(2)}(0)$ when the visibility is extracted from the \textit{intensity correlator} \cite{ollivier2021hong}. 
\end{itemize}
The remainder of this work is dedicated to deriving and substantiating these results.

\section{Imperfect single-photon state model}
We begin by modeling the quantum state emitted by an imperfect SPS, starting with a brief review of the ideal state produced by a perfect SPS. An ideal SPS generates on-demand single excitations in a well-defined \textit{external} mode $i$ (e.g., spatial) of the quantized electromagnetic field, producing indistinguishable single-photon states that are identical in all other measurable \textit{internal} modes (e.g., polarization, frequency, and arrival time). The perfect photon state is formally expressed as $\ket{\psi} := \hat{a}_i^\dagger\ket{0} = \ket{1}_i$. Here, $\hat{a}_i^\dagger = \sum_{n \geq 1} \sqrt{n}\ket{n}\bra{n-1}_i$ is the bosonic creation operator that acts on the vacuum state $\ket{0}$ to produce an ideal single-photon Fock state $\ket{1}$ in mode $i$. However, both coherent and incoherent noise processes can excite undesired internal modes, producing photons that are partially distinguishable from the ideal state.

Throughout this work, we assume that excitations in internal modes cannot be individually resolved or controlled. Under this assumption, the noisy excitation is described by the partially distinguishable photon state $\rho(\epsilon)$, where $\epsilon$ quantifies the indistinguishability error via wavefunction overlap with the ideal state \cite{marshall2022distillation}:
\begin{equation}
    \epsilon:= 1- \langle 1 | \rho(\epsilon) | 1 \rangle. 
\end{equation}

\begin{figure}
    \centering
     \includegraphics[width=8.6cm, height=20cm, keepaspectratio,]{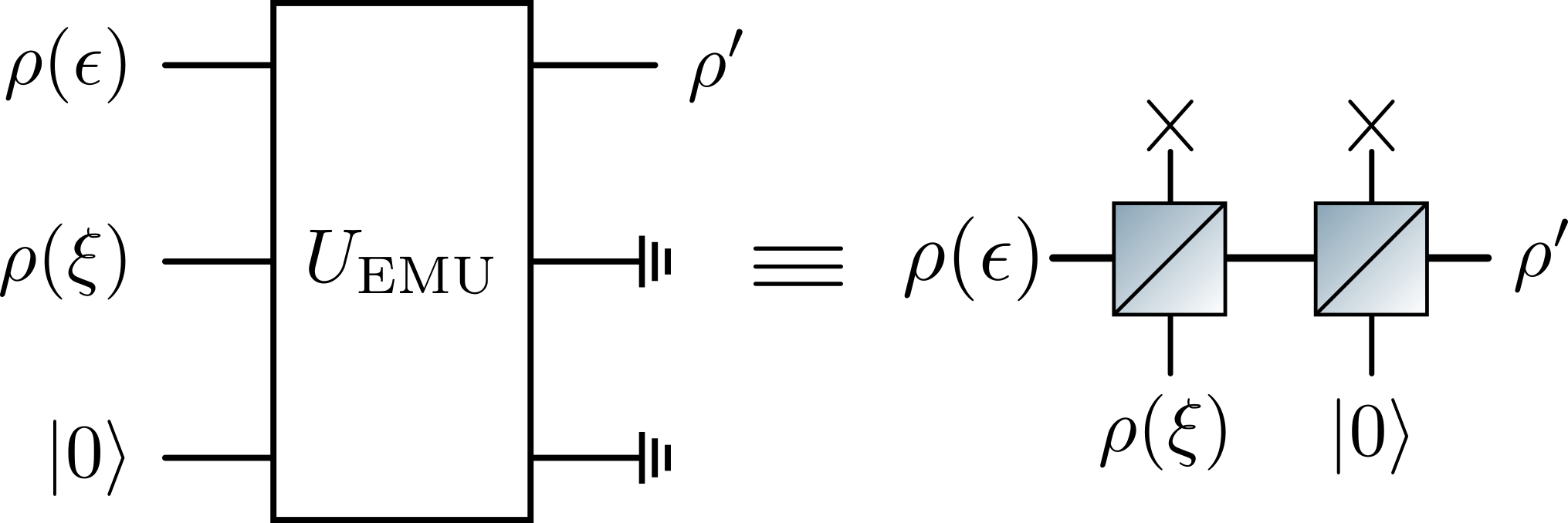}
    \caption{\textbf{Linear optical circuit for emulating imperfect single-photon states.} 
    A partially distinguishable signal photon is interfered with a partially distinguishable noise photon and vacuum to emulate indistinguishability errors, photon-number impurity, and loss errors. The process is modeled as a sequence of beam splitter interactions: the first mixes the signal photon $\rho(\epsilon)$ with the noise photon $\rho(\xi)$, introducing multiphoton noise; the second couples in vacuum $\ket{0}$ to model loss. The combined effect defines a purification of the quantum channel acting on the input state $\rho(\epsilon)\otimes\rho(\xi)\otimes\ket{0}\bra{0}$, described by a unitary transformation matrix $U_{\text{EMU}}$. The resulting reduced state $\rho'$ defines the imperfect single-photon state model used throughout this work.}
    \label{fig:EMU}
\end{figure}

We now study the propagation of such partially distinguishable photon states through a specific linear optical circuit designed to emulate additional imperfections, specifically photon-number impurity and loss (Fig. \ref{fig:EMU}). The evolution is described by a unitary matrix $U_{\text{EMU}}$ that governs the transformation of the creation operators, where we assume that the transformation $ \hat{a}_i^\dagger \to \sum_{j=1} (U_{\text{EMU}})_{ij} \hat{a}_j^\dagger$ is agnostic to the specific internal mode structure of the excitation. In our model, we implicitly assume that multiphoton contamination of the emitted photon state is limited to at most two-photon components, which is commonly assumed for a first-order correction in photon-number impurity. We model the photon-number impurity error as a beam splitter interaction between a signal photon $\rho(\epsilon)$ and a noise photon $\rho(\xi)$ \cite{ollivier2021hong}. In contrast to our approach, alternative models omit this agnostic treatment of signal and noise photons \cite{pont2024high}. Nevertheless, this does not affect the reported key findings (see \textit{Discussion}). The loss error is modeled by the second beam splitter interaction, where the monitored output mode is coupled to an auxiliary vacuum mode \cite{oszmaniec2018classical}. For our purposes, the unitary transformation matrix of an ideal beam splitter is fully parametrized by the reflectivity, $R$: 
\begin{equation}
    U_{\text{BS}} = \begin{bmatrix}
    \sqrt{R} & \sqrt{1-R} \\
    \sqrt{1-R} & -\sqrt{R}
    \end{bmatrix}. 
    \label{eq:U_BS}
\end{equation}
First, we parametrize the signal–noise photon interaction using a multiphoton error parameter $p$, such that the reflectivity of the corresponding beam splitter is given by $R = p$. In the limiting cases, $p = 0$ corresponds to no multiphoton contamination, while $p = \frac{1}{2}$ corresponds to significant photon-number impurity errors. These serve as illustrative examples of the parameter’s physical meaning, which we will later connect to the measurable quantum observable $g^{(2)}(0)$.  Second, we parametrize the signal-vacuum interaction using a transmission efficiency parameter $\eta$. The loss is thus quantified by $1- \eta$, corresponding to a beam splitter reflectivity of $R = 1 - \eta$. In summary, the equivalent concatenated beam splitter interaction shown in Fig. \ref{fig:EMU} (modulo permutations of input and output ports) is given by:
\begin{equation}
    U_{\text{EMU}} = 
    \begin{bsmallmatrix}
    \sqrt{\eta(1-p)} & \sqrt{p} & \sqrt{(1-\eta)(1-p)} \\
    \sqrt{\eta p} & -\sqrt{1-p} & \sqrt{(1-\eta)p} \\
    \sqrt{1-\eta} & 0 & -\sqrt{\eta}
    \end{bsmallmatrix}.
    \label{eq:U_EMU}
\end{equation}

Finally, we propagate $\rho(\epsilon)\otimes\rho(\xi)\otimes\ket{0}\bra{0}$ through the emulator circuit shown in Fig. \ref{fig:EMU} to derive the imperfect single-photon state model $\rho^\prime$ studied in this work:
\begin{equation}
    \rho^\prime = \Phi[\rho(\epsilon)\otimes\rho(\xi)\otimes\ket{0}\bra{0}]. 
\end{equation}
Here, $\Phi[\cdot]$ denotes the quantum operation implemented by the emulator circuit, which transforms the creation operators according to the unitary $U_{\text{EMU}}$ and traces out the virtual modes. This imperfect state can be conveniently represented as a statistical mixture of photon number states:
\begin{equation}
    \rho^\prime = P_0\ket{0}\bra{0} + P_1\rho_1 + P_2 \rho_2, 
    \label{eq:ISPSM}
\end{equation}
where $P_n$ denotes the probability associated with the $n$-photon number state $\rho_n$. The key characteristics of this state model are quantified by the single-photon trace purity, defined as $\Tr[\rho_1^2]$, and second-order correlation function, defined as:
\begin{equation}
    g^{(2)}(0) = \frac{2P_2}{(P_1 + 2P_2)^2}.
    \label{eq:def_g2}
\end{equation}

So far, we have not explicitly discussed the form of partially distinguishable photons \cite{annoni2025incoherent}. In general, the internal mode structure of such photons can be complex, with coherent partial overlaps that give rise to rich multiphoton interference phenomena \cite{menssen2017distinguishability,seron2023boson}. In this work, we simplify the error analysis by modeling each partially distinguishable photon $k$ using the \textit{orthogonal bad bit} (OBB) model \cite{sparrow2017quantum}:
\begin{equation}
    \rho_k(\epsilon) = (\sqrt{1-\epsilon}\ket{1} + \sqrt{\epsilon}\ket{\tilde{1}_k})(\sqrt{1-\epsilon}\bra{1} + \sqrt{\epsilon}\bra{\tilde{1}_k}),
\end{equation}
where $\ket{\tilde{1}_k}$ denotes an orthogonal error mode satisfying $\braket{1}{\tilde{1}_k} = 0$ and $\braket{\tilde{1}_k}{\tilde{1}_l} = \delta_{kl}$.  In this model, indistinguishability errors are treated as flips to orthogonal internal modes that are not experimentally resolvable. We therefore omit the subscript $k$ on the error modes throughout this work. Consequently, each partially distinguishable photon can be equivalently described by the mixed state:
\begin{equation}
    \rho(\epsilon) = (1-\epsilon)\ket{1}\bra{1} + \epsilon \ket{\tilde{1}}\bra{\tilde{1}}, 
    \label{eq:rho_1_before_loss}
\end{equation}
since coherence between $\ket{1}$ and $\ket{\tilde{1}}$ is not accessible \cite{annoni2025incoherent}. Under this model, the intrinsic overlap of two signal photons is simply given by $\Tr[\rho(\epsilon)^2 ] = (1-\epsilon)^2$. 

In our model, the indistinguishability error of the noise photon $\xi$ is treated as an independent parameter to phenomenologically account for different multiphoton noise processes, with literature typically reporting extreme cases: $\xi = 1$ for fully distinguishable noise photons, as in quantum dots \cite{ollivier2021hong, tomm2021bright}, or $\xi = \epsilon$ for noise photons identical to the signal photons, as in heralded parametric down-conversion sources \cite{tsujimoto2021ultra,tsujimoto2023quantum} (see also Appendix \ref{app:identicalcorrelatednoise}). By allowing $\xi$ to vary, we aim to capture an arbitrary combination of multiphoton noise processes, in which a combination of identical and distinguishable noise photons contribute to the imperfect single-photon state, motivating the parameter restriction $0 \leq \epsilon \leq \xi \leq 1$.

While the OBB model represents a significant simplification of how partial distinguishability errors manifest in photonic systems, its use is justified by our focus on the high-visibility regime \cite{renema2021sample}. In this regime, the parameter $\epsilon$ is by definition small, and an expansion of a product state of $N$ OBB photons shows that the leading-order error approximation corresponds to a state in which $N-1$ photons remain in the target mode, while one photon occupies an orthogonal error mode \cite{moylett2019classically}. A similar leading-order structure appears in the most conceptually distinct alternative, the \textit{same bad bit} (SBB) model \cite{saied2024advancing}, in which all erroneous photons occupy a common orthogonal error mode. Although the validity of these first-order approximations diminishes with increasing photon number $N$ \cite{somhorst2025photon}, we note that most photonic quantum gates involve interference among only a small number of photons \cite{chen2024heralded,cao2024photonic,maring2024versatile,bayerbach2023bell,guo2024boosted,hauser2025boosted}. This supports the applicability of the OBB model in many practical scenarios.

We proceed by deriving expressions for the model parameters in Eq. \ref{eq:ISPSM}. First, we decompose the input state into a product of distinguishable and indistinguishable photon contributions \cite{moylett2019classically}:
\begin{equation}
    \begin{split}
      \rho(\epsilon)\otimes\rho(\xi)\otimes\ketbra{0}{0} &=(1-\epsilon)(1-\xi)\ketbra{110}{110} \\&+(1-\epsilon)\xi \ketbra{1\tilde{1}0}{1\tilde{1}0} \\&+ \epsilon(1-\xi)\ketbra{\tilde{1}10}{\tilde{1}10}\\&+\epsilon \xi \ketbra{\tilde{1}\tilde{1}0}{\tilde{1}\tilde{1}0}.
    \end{split}
\end{equation}
Next, we determine how each term contributes to the emulated imperfect single-photon state: 
\begin{equation}
    \begin{split}
        \Phi[\ketbra{110}{110}] & = (1-\eta + 2\eta^2p)\ketbra{0}{0} \\ &+ \eta(1-4\eta p)\ketbra{1}{1} \\ & + 2\eta^2p\ketbra{2}{2}, 
    \end{split}
\end{equation}
\begin{equation}
    \begin{split}
        \Phi[\ketbra{1\tilde{1}0}{1\tilde{1}0}] &= (1-\eta + \eta^2p)\ketbra{0}{0} \\&+(\eta(1-p) -\eta^2 p)\ketbra{1}{1} \\&+\eta(1-\eta)p\ketbra{\tilde{1}}{\tilde{1}} \\&+\eta^2p\ketbra{(1,\tilde{1})}{(1,\tilde{1})}, 
    \end{split}
\end{equation}
\begin{equation}
    \begin{split}
        \Phi[\ketbra{\tilde{1}10}{\tilde{1}10}] & = (1-\eta + \eta^2p)\ketbra{0}{0} \\ & + \eta(1-\eta)p\ketbra{1}{1} \\ & +  (\eta(1-p) -\eta^2 p)\ketbra{\tilde{1}}{\tilde{1}} \\&+ \eta^2p\ketbra{(1,\tilde{1})}{(1,\tilde{1})}, 
    \end{split}
\end{equation}
\begin{equation}
    \begin{split}
        \Phi[\ketbra{\tilde{1}\tilde{1}0}{\tilde{1}\tilde{1}0}] &=(1-\eta + \eta^2p)\ketbra{0}{0} \\&+\eta(1-2\eta p)\ketbra{\tilde{1}}{\tilde{1}} \\&+\eta^2p\ketbra{(\tilde{1},\tilde{1})}{(\tilde{1},\tilde{1})}. 
    \end{split}
\end{equation}
Grouping terms by total photon number, we find $P_0 = 1 - \eta + \eta^2p(1+(1-\epsilon)(1-\xi))$, $P_2 = \eta^2 p (1+(1-\epsilon)(1-\xi))$, and consequently, $P_1 = \eta - 2\eta^2p(1+(1-\epsilon)(1-\xi))$. Based on this photon-number probability distribution and Eq. \ref{eq:def_g2}, we derive a theoretical expression for the experimentally accessible second-order correlation function:
\begin{equation}
    g^{(2)}(0) = 2p(1+(1-\epsilon)(1-\xi)). 
\end{equation}
This relationship can also be inverted: the photon-number probabilities in Eq. \ref{eq:ISPSM} can be expressed as functions of the measurable $g^{(2)}(0)$:
\begin{itemize}
    \item $P_0 = 1 - \eta + \frac{1}{2}\eta^2 g^{(2)}(0)$,
    \item $P_1 = \eta  - \eta^2 g^{(2)}(0)$, 
    \item $P_2 = \frac{1}{2}\eta^2 g^{(2)}(0)$. 
\end{itemize}

We now turn to the single-photon trace purity $\Tr[\rho_1^2]$. In general, $\Tr[\rho_1^2] \neq  (1-\epsilon)^2$, as the combined effects of optical loss and multiphoton noise allow the noise-photon state $\rho(\xi)$ to leak into the single-photon component \cite{ollivier2021hong,gonzalez2025two}. To account for this, we model the effective single-photon state as $\rho_1 = \rho(\tilde{\epsilon})$, where $\tilde{\epsilon}$ represents an \textit{effective} indistinguishability error that incorporates both loss and multiphoton contributions. After collecting terms and simplifying, we obtain the following condition necessary for consistency:
\begin{equation}
    P_1 \tilde{\epsilon} = \eta \left(\epsilon + \frac{1}{2}\frac{\xi - \epsilon - \eta(\xi + \epsilon)}{1 + (1-\epsilon)(1-\xi)} g^{(2)}(0)  \right), 
\end{equation}
which leads to the following expression for the effective indistinguishability error:
\begin{equation}
    \begin{split}
    \tilde{\epsilon} & = \epsilon + \frac{1}{2}(1-\eta)\frac{\xi - \epsilon}{1 + (1-\epsilon)(1-\xi)}g^{(2)}(0) \\ & + \eta \frac{\epsilon(1-\epsilon)(1-\xi)}{1+(1-\epsilon)(1-\xi)}g^{(2)}(0) + \mathcal{O}\left( g^{(2)}(0)^2 \right).
    \end{split}
    \label{eq:eps_eff}
\end{equation}
Our analysis reveals that the effective indistinguishability error is lower bounded by the intrinsic indistinguishability error, $\epsilon \leq \tilde{\epsilon}$, which has important implications for benchmarking practical SPSs that are also subject to loss and photon-number purity imperfections. 
Analogous to the intrinsic single-photon trace purity, the purity in the presence of such imperfections is given by:
\begin{equation}
    \Tr[\rho_1^2] = (1-\tilde{\epsilon})^2.
\end{equation}

To summarize, in this section we analyzed a linear optical circuit that emulates an imperfect photon state, incorporating loss, partial distinguishability, and multiphoton errors. We derived expressions for the key metrics $g^{(2)}(0)$ and $\Tr[\rho_1^2]$. In particular, we verified that the combination of loss and multiphoton effects reduces the trace purity of the single-photon component. To capture these effects, we introduced the concept of an \textit{effective} indistinguishability error. In the next section, we use this model to interpret the results of HOM interference experiments.

\section{Derivation of correction factors}
In the previous section, we derived expressions for the characteristics $\Tr[\rho_1^2]$ and $g^{(2)}(0)$ based on an imperfect single-photon model. In this section, we apply the model to experimentally relevant quantities: the measured HOM correlator $g_{\text{HOM}}$ and the measured HOM visibility $V_{\text{HOM}}$ \cite{trivedi2020generation}. By relating these observables to the underlying imperfect photon model, we develop measurement-dependent correction factors that enable the extraction of the effective indistinguishability error from experimental data.

\begin{figure}[h!]
    \centering
     \includegraphics[width=4cm, height=20cm, keepaspectratio,]{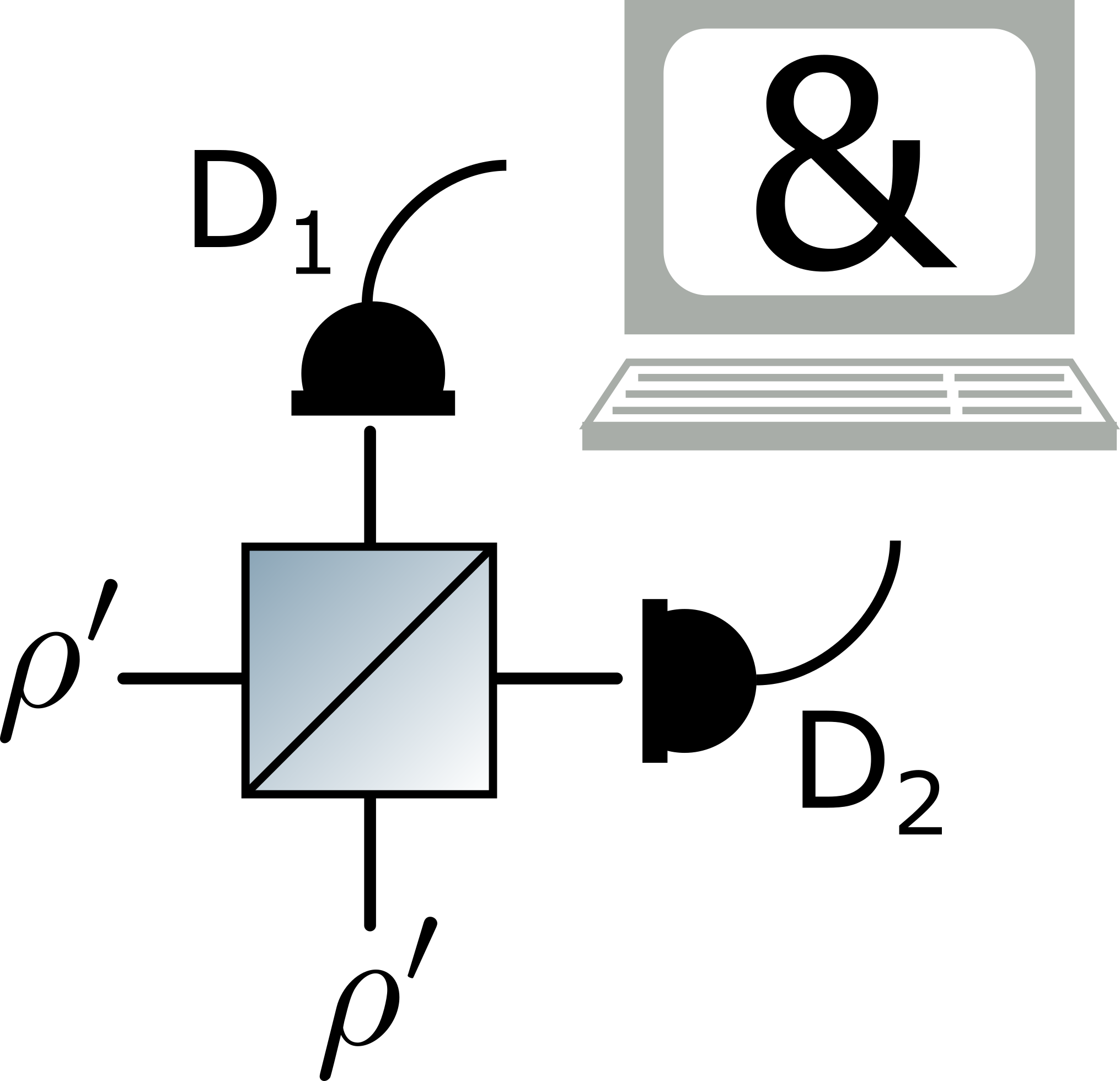}
    \caption{\textbf{Setup for a Hong-Ou-Mandel interference measurement with imperfect single-photon states.} Two copies of an imperfect photon state $\rho^\prime$ are interfered on a perfectly balanced beam splitter ($R = \frac{1}{2}$) and detected by click/no-click detectors $D_1$ and $D_2$. An electronic correlator is used to extract time-correlated click events between the two detectors within a narrow coincidence window, while also recording marginal click events over a set integration time.  In certain measurement protocols, a reference value is obtained by intentionally making one copy orthogonal to its counterpart and repeating the interference experiment.}
    \label{fig:HOM}
\end{figure}

We begin by revisiting the basic principles of the HOM interference experiment \cite{hong1987measurement}. In a typical setup, two photons are interfered at a balanced two-port beam splitter  ($R = \frac{1}{2}$, Eq. \ref{eq:U_BS}), followed by detection in the photon number basis, usually with detectors lacking number-resolving capability (Fig. \ref{fig:HOM}). The key signature of HOM interference is photon bunching in the output statistics: the more indistinguishable the photons, the less likely they are to exit the beam splitter separately. Experimentally, this is often quantified by measuring the coincidence rate for partially distinguishable photons and comparing it to the measured rate for fully distinguishable photons ($\tilde{\epsilon} = 1$), typically achieved by introducing a time delay or polarization rotation, hereafter referred to as \textit{Method A}:
\begin{equation}
    V^{A}_{\text{HOM}} = 1 - \frac{P(D_1 \cap D_2)}{P(D_1 \cap D_2 | \tilde{\epsilon} = 1)},
    \label{eq:V_raw_A}
\end{equation}
where $P(D_1 \cap D_2)$ is the joint detection probability at detectors 1 and 2, proportional to the measured coincidence rate. 

If preparing a fully distinguishable reference state is not feasible, the HOM correlator $g^{(2)}_{\text{HOM}}(0)$ is commonly used instead (referred to as \textit{Method B}) \cite{trivedi2020generation}. For convenience, we abbreviate this quantity as $g_{\text{HOM}}$. $g_{\text{HOM}}$ is analogous to $g^{(2)}(0)$, which characterizes the normalized autocorrelation of a single extrinsic mode, but here the normalized cross-correlation is used to probe two-photon coherence \cite{fischer2016dynamical}:
\begin{equation}
    g_{\text{HOM}} := \frac{P(D_1 \cap D_2)}{P(D_1) P(D_2)},
    \label{eq:g_HOM}
\end{equation}  
where $P(D_i)$ is the marginal click probability of detector $i$. Finally, the visibility is approximated using the relation \cite{ollivier2021hong}:
\begin{equation}
    V^{B}_{\text{HOM}} = 1 - 2g_{\text{HOM}}.
\end{equation}

We now apply our model to interpret the experimentally measurable $g_{\text{HOM}}$ and $V_{\text{HOM}}$. Under typical experimental conditions of low transmittance ($\eta \approx 0$) and small multiphoton contributions ($g^{(2)}(0) \approx 0$), the input state for the HOM experiment (Fig. \ref{fig:HOM}) can be approximated as:
\begin{equation}
    \begin{split}
        \rho^\prime &\otimes \rho^\prime =  \left(1 - 2\eta + \eta^2 + \eta^2 g^{(2)}(0) \right)\ketbra{00}{00} \\
        & + \left( \eta - \eta^2 - \eta^2 g^{(2)}(0)\right) \left(\ketbra{0}{0}\otimes \rho(\tilde{\epsilon}) + \rho(\tilde{\epsilon})\otimes\ketbra{0}{0} \right) \\
        & + \frac{1}{2}\eta^2 g^{(2)}(0) \left( \ketbra{0}{0}\otimes \rho_2 + \rho_2\otimes\ketbra{0}{0} \right) \\
        & + \eta^2 \rho(\tilde{\epsilon})\otimes\rho(\tilde{\epsilon}) + \mathcal{O}\left(\eta^3 g^{(2)}(0) \right) + \mathcal{O}\left(\eta^4 (g^{(2)}(0))^2 \right). 
        \label{eq:rhotimesrho}
    \end{split}
\end{equation}
After the balanced beam splitter interaction, the detection click probabilities are:
\begin{equation}
    \begin{split}
        P(D_1\cap D_2) & = \frac{1}{2}\eta^2g^{(2)}(0) \left( \frac{1}{2}+\frac{1}{2} \right) + \eta^2 \frac{1}{2}(1-(1-\tilde{\epsilon})^2) \\
        & = \frac{1}{2}\eta^2(1-(1-\tilde{\epsilon})^2 + g^{(2)}(0)),
    \end{split}
\end{equation}

\begin{equation}
    \begin{split}
        P(D_i) & = \left(\eta-\eta^2 - \eta^2 g^{(2)}(0) \right)\left(\frac{1}{2} + \frac{1}{2} \right) \\ & + \frac{1}{2}\eta^2g^{(2)}(0)\left(\frac{3}{4} + \frac{3}{4}\right) + \eta^2 \left( \frac{3}{4} - \frac{1}{4}(1-\tilde{\epsilon})^2 \right)\\
        & = \eta - \frac{1}{4} \eta^2 \left(1 + (1-\tilde{\epsilon})^2 + g^{(2)}(0) \right), 
    \end{split}
\end{equation}
such that:
\begin{equation}
    \begin{split}
        P(D_1)P(D_2) & = \eta^2 - \frac{1}{2}\eta^3(1+(1-\tilde{\epsilon})^2) \\ & + \mathcal{O}\left(\eta^3 g^{(2)}(0) \right) + \mathcal{O}\left(\eta^4 (g^{(2)}(0))^2 \right).
    \end{split}
\end{equation}
From this point forward, every equation is up to first order in $\eta$ and $g^{(2)}(0)$, unless otherwise specified. Therefore, the HOM correlator can be approximated as:
\begin{equation}
    g_{\text{HOM}} = \frac{1}{2}\left( 1 - (1-\tilde{\epsilon})^2 + g^{(2)}(0) + \frac{1}{2}\eta \left(1 - (1-\tilde{\epsilon})^4 \right) \right). 
\end{equation}

Using the earlier expressions, we interpret the raw visibilities as follows:
\begin{equation}
    V^{A}_{\text{HOM}} = \frac{(1-\tilde{\epsilon})^2}{1 + g^{(2)}(0)},
    \label{eq_V_A_eff}
\end{equation}
and
\begin{equation}
    V^{B}_{\text{HOM}} = (1-\tilde{\epsilon})^2 - g^{(2)}(0) - \frac{1}{2}\eta(1-(1-\tilde{\epsilon})^4). 
    \label{eq:V_B_eff}
\end{equation}

The first-order approximation in the transmission efficiency remains accurate provided that the experimental conditions satisfy $\eta \leq g^{(2)}(0)$. This condition is justified because we used the standard multiphoton assumption $P_{n>2} = 0$ to derive our model (Fig. \ref{fig:EMU}), which is valid when higher-order terms $\mathcal{O}\left( (g^{(2)}(0))^2 \right)$ can be neglected \footnote{Ref. \cite{ollivier2021hong} reports that the approximation $\mathcal{O}\left( (g^{(2)}(0))^2 \right) \approx 0$ holds well for $ g^{(2)}(0))^2 < 0.3$.}. In Eq. \ref{eq:rhotimesrho}, the higher-order efficiency error terms scale as $\mathcal{O}\left( \eta^3 g^{(2)}(0) \right)$ and $\mathcal{O}\left( \eta^4 (g^{(2)}(0))^2 \right)$, with only the first term contributing significantly under our multiphoton error modeling assumptions. In the visibility correction formulas, this first term contributes a leading error of order $\mathcal{O}\left( \eta g^{(2)}(0) \right)$. Hence, the simplified model is accurate under the adopted approximations as long as $\eta g^{(2)}(0) \leq (g^{(2)}(0))^2$, yielding the restriction $\eta \leq g^{(2)}(0)$. To accurately describe the high-efficiency transmission regime, the model of $\rho^\prime \otimes \rho^\prime$ must be extended to include three-photon interference terms between input state components $\rho_2 \otimes \rho_1$ and $\rho_1 \otimes \rho_2$, which contribute at $\mathcal{O}\left( \eta^3 g^{(2)}(0) \right)$. We leave this refinement for future work. 

To match the level of approximation commonly used in literature, we neglect all first-order corrections in $\eta$ from here onward. Eqs. \ref{eq_V_A_eff} and \ref{eq:V_B_eff} yield measurement-dependent visibility correction factors necessary to estimate the effective photonic indistinguishability error $\tilde{\epsilon}$:
\begin{equation}
    (1-\tilde{\epsilon})^2 = V^{A}_{\text{HOM}} \left( 1 + g^{(2)}(0) \right), 
\end{equation}
and
\begin{equation}
    (1-\tilde{\epsilon})^2 = V^{B}_{\text{HOM}} + g^{(2)}(0).
\end{equation}
However, to estimate the intrinsic indistinguishability error, additional information about the indistinguishability of the noise photons is required. As argued in Ref. \cite{ollivier2021hong}, it is often justified to assume $\xi = 1$ for common quantum dot SPSs. Under this assumption, $\tilde{\epsilon} = \epsilon + \frac{1}{2}(1-\epsilon)g^{(2)}(0)$. For Method A, this yields:
\begin{equation}
    (1-\epsilon)^2 = V^{A}_{\text{HOM}} \left( 1 + 2g^{(2)}(0) \right),
    \label{eq:eps_A}
\end{equation}
consistent with prior results \cite{tomm2021bright}. 

Similarly, for Method B, using first the unspecified substitution $\tilde{\epsilon} = \epsilon + \frac{1}{2} \frac{\xi - \epsilon}{1 + (1-\epsilon)(1-\xi)} g^{(2)}(0)$, the expression becomes:
\begin{equation}
    V^{B}_{\text{HOM}} = (1 - \epsilon)^2 - \left( \frac{1 + (1-\epsilon)^2}{1 + (1-\epsilon)(1-\xi)} \right) g^{(2)}(0),
\end{equation}
which is consistent with the main result of Ref. \cite{ollivier2021hong}. Substituting $\xi = 1$ yields consistently:
\begin{equation}
    (1-\epsilon)^2 = \frac{V^{B}_{\text{HOM}} + g^{(2)}(0)}{1 - g^{(2)}(0)}.
    \label{eq:eps_B}
\end{equation}
These examples support the validity of our work. 

To summarize, in this section we reviewed the HOM interference experiment, where visibility is typically measured using Method A (coincidence counts) or Method B (intensity correlator). We identified a measurement-dependent correction factor that accounts for photon-number impurity when estimating the effective indistinguishability error. Our method yields intrinsic indistinguishability errors that are consistent with expressions reported in the literature.

\section{Discussion}
In summary, we presented a method to extract the photonic indistinguishability error from measurable quantum observables. 
First, we introduced a general model for imperfect single-photon states that incorporates realistic experimental errors, including loss, partial distinguishability, and photon-number impurity. Building on this model, we defined the concept of an \textit{effective} indistinguishability error, which is generally larger than the \textit{intrinsic} error due to the combined effects of loss and multiphoton noise. We propose this effective error as a more practical measure for evaluating single-photon interference performance, with potential impact on the design and analysis of photonic quantum information protocols. Finally, we demonstrated how this parameter can be extracted from experimentally measurable observables, specifically $g^{(2)}(0)$ and $V_{\text{HOM}}$, with the visibility correction factor shown to depend on the chosen measurement protocol. 

Beyond applications in single-photon source characterization, our approach can improve photonic quantum information protocols that rely on genuine single-photon overlap measurements in a specific degree of freedom, such as quantum state tomography \cite{white2025robust} and quantum network diagnostics \cite{valivarthi2020teleportation,basso2019entanglement}. When an estimate of the multiphoton indistinguishability is available, our method can help mitigate the effects of loss and multiphoton noise through postselection, enabling more accurate extraction of the underlying single-photon trace purity under realistic experimental conditions.

While our analysis provides valuable insights for a wide range of SPSs, certain limitations to its applicability remain. First, we emphasize that the scope of this work is restricted to separable multiphoton noise. Consequently, the applicability of our model to more complex sources involving entangled noise photons remains an open question. For example, it may not fully capture the behavior of imperfect photon states produced by heralded parametric down-conversion sources, where the internal mode of the noise photon often overlaps significantly with that of the signal photon due to correlations introduced by the squeezing process \cite{eckstein2011highly}. Although adjusting $\xi$ helps to address this problem, our model may not fully capture the bunching effects arising from photon-number entanglement, which enhance the two-photon component. However, recent experiments support the use of a separate noise model, where $\xi = \epsilon$ \cite{tsujimoto2021ultra, tsujimoto2023quantum}. In Appendix \ref{app:identicalcorrelatednoise}, we verify that correlated identical noise has the same effect on measured visibilities as uncorrelated identical noise.
Second, our model provides a continuously tunable parametrization of the noise photon indistinguishability error. In the experimentally relevant case of fully distinguishable noise photons, an alternative model may be considered in which the first beam splitter no longer treats the signal and noise photons identically \cite{pont2024high}. Conceptually, this resembles the action of a dichroic mirror or polarizing beam splitter that merges two photons with distinct internal modes into a single extrinsic output mode. However, the trace distance $\delta$ \cite{nielsen2010quantum} between the two models is typically small under experimentally relevant conditions, $\delta = \mathcal{O} \left( \eta g^{(2)}(0) \right)$, supporting the general validity of our model \footnote{In fact, these small differences result in $P_0 = 1 - \eta - \frac{1}{2}\eta g^{(2)}(0) + \frac{1}{2}\eta^2 g^{(2)}(0)$ and $P_1 = \eta + \frac{1}{2}\eta g^{(2)}(0) - \eta^2 g^{(2)}(0)$, but they do not affect the expression for the effective indistinguishability error (Eq. \ref{eq:eps_eff}) nor the expressions for the correction formulas (Eqs. \ref{eq:eps_A} and \ref{eq:eps_B}).}.

\section{Competing interests}
The authors declare that they have no conflict of interest.\\

\section{Acknowledgements}
F.H.B.S. acknowledges the AQO-QIP and NASA QuAIL teams for their valuable scientific discussions. This research is supported by the Photonic Integrated Technology Center (PITC). This publication is part of the project ``At the Quantum Edge" of the research programme VIDI, which is financed by the Dutch Research Council (NWO). 
J.S. and E.G.R. are grateful for support from  DARPA under IAA 8839, Annex 130, and from  NASA Ames Research Center. 
The United States Government retains, and by accepting the article for publication, the publisher acknowledges that the United States Government retains, a nonexclusive, paid-up, irrevocable, worldwide license to publish or reproduce the published form of this work, or allow others to do so, for United States Government purposes.

\appendix
\section{Applicability to heralded parametric down-conversion sources}
\label{app:identicalcorrelatednoise}
In this Appendix, we examine the accuracy of the uncorrelated noise model presented in the main text for describing widely used SPSs  subject to correlated identical noise. As discussed in the main text, heralded SPSs based on parametric down-conversion are prototypical examples of such systems. 

In the absence of photon indistinguishability errors and loss, the two-mode state generated by a parametric down-conversion source can be written as \cite{eckstein2011highly}:
\begin{equation}
    \ket{\Psi} = \sqrt{1-p} \sum_{n=0}^{\infty} \sqrt{p}^n \ket{n}\ket{n}. 
\end{equation}
When operated as a heralded SPS, a projective measurement is performed on one mode, typically using a non–number-resolving detector. A detection event heralds that there are $n \geq 1$ photons in the unmeasured mode due to the photon-number correlations, thereby heralding an imperfect single-photon state described by the reduced density operator:
\begin{equation}
    \begin{split}
    \sigma_{h} &= \sum_{n=1}^{\infty}(p^{n-1}-p^n)\ket{n}\bra{n} \\
    &= (1-p)\ket{1}\bra{1} + p\ket{2}\bra{2} + \mathcal{O}\left( p^2 \right). 
    \end{split}
\end{equation}
This state serves as the basis for the phenomenological model of heralded SPSs considered here and corresponds to the intermediate state before losses, e.g., the state between the beam splitters in Fig. \ref{fig:EMU}:
\begin{equation}
    \rho_{h} = (1-p)\rho_1 + p \rho_2, 
\end{equation}
where $\rho_1$ is given by Eq. \ref{eq:rho_1_before_loss}, and the two-photon component is given by:
\begin{equation}
    \rho_2 = (1-\epsilon)\ket{2}\bra{2} + \epsilon\ket{\tilde{2}}\bra{\tilde{2}}. 
\end{equation}
This two-photon component satisfies the property that tracing out one photon leaves the single-photon state of Eq. \ref{eq:rho_1_before_loss}, i.e., it exhibits correlated identical noise. Hence, we compare $\rho_{h}$ with the uncorrelated identical noise model ($\xi = \epsilon$). 

When the effects of loss are included through the standard beam-splitter model, the resulting imperfect heralded single-photon state takes the same form as Eq. \ref{eq:ISPSM}. However, in this case, the photon-number probabilities slightly differ from those in the uncorrelated identical noise model:
\begin{itemize}
    \item $P_0 = 1 - \eta - \frac{1}{2}\eta(1-\eta)g^{(2)}(0)$,
    \item $P_1 = \eta  +\frac{1}{2}\eta(1-2\eta)g^{(2)}(0)$, 
    \item $P_2 = \frac{1}{2}\eta^2 g^{(2)}(0)$,
\end{itemize}
where $g^{(2)}(0) = 2p + \mathcal{O}\left( p^2 \right)$.

Using these probabilities, we obtain:
\begin{equation}
    \begin{split}
        \rho_h^\prime &\otimes \rho_h^\prime =  \left(1 - 2\eta + \eta^2 - \eta g^{(2)}(0) + 2\eta^2 g^{(2)}(0)  \right)\ketbra{00}{00} \\
        & + \left( \eta - \eta^2 + \frac{1}{2}\eta g^{(2)}(0) - 2\eta^2 g^{(2)}(0) \right) \ketbra{0}{0}\otimes \rho(\epsilon) \\&
        + \left( \eta - \eta^2 + \frac{1}{2}\eta g^{(2)}(0) - 2\eta^2 g^{(2)}(0) \right)\rho(\epsilon)\otimes\ketbra{0}{0}\\
        & + \frac{1}{2}\eta^2 g^{(2)}(0) \left( \ketbra{0}{0}\otimes \rho_2 + \rho_2\otimes\ketbra{0}{0} \right) \\
        & + \left( \eta^2 + \eta^2 g^{(2)}(0) \right)\rho(\epsilon)\otimes\rho(\epsilon) \\& + \mathcal{O}\left(\eta^3 g^{(2)}(0) \right) + \mathcal{O}\left(\eta^4 (g^{(2)}(0))^2 \right). 
    \end{split}
\end{equation}

After the balanced beam splitter interaction, the detection
click probabilities are:
\begin{equation}
    \begin{split}
        P(D_1\cap D_2) = \frac{1}{2}\eta^2 \left( ( 1 - (1-\epsilon)^2 )(1 + g^{(2)}(0) ) + g^{(2)}(0) \right),
    \end{split}
\end{equation}
and,
\begin{equation}
    \begin{split}
        P(D_1)P(D_2) & = \eta^2(1 + g^{(2)}(0) ) - \frac{1}{2}\eta^3(1+(1-\epsilon)^2) \\ & + \mathcal{O}\left(\eta^3 g^{(2)}(0) \right) + \mathcal{O}\left(\eta^4 (g^{(2)}(0))^2 \right).
    \end{split}
\end{equation}
It can be shown that the resulting visibilities coincide with those reported in the main text, provided that $\tilde{\epsilon} = \epsilon$. This indicates that the phenomenological correlated identical noise model is effectively equivalent to the uncorrelated identical noise model. The applicability of this conclusion is supported by recent experimental results \cite{tsujimoto2021ultra,tsujimoto2023quantum}.

\bibliography{refs.bib}

\begin{thebibliography}{49}%
\makeatletter
\providecommand \@ifxundefined [1]{%
 \@ifx{#1\undefined}
}%
\providecommand \@ifnum [1]{%
 \ifnum #1\expandafter \@firstoftwo
 \else \expandafter \@secondoftwo
 \fi
}%
\providecommand \@ifx [1]{%
 \ifx #1\expandafter \@firstoftwo
 \else \expandafter \@secondoftwo
 \fi
}%
\providecommand \natexlab [1]{#1}%
\providecommand \enquote  [1]{``#1''}%
\providecommand \bibnamefont  [1]{#1}%
\providecommand \bibfnamefont [1]{#1}%
\providecommand \citenamefont [1]{#1}%
\providecommand \href@noop [0]{\@secondoftwo}%
\providecommand \href [0]{\begingroup \@sanitize@url \@href}%
\providecommand \@href[1]{\@@startlink{#1}\@@href}%
\providecommand \@@href[1]{\endgroup#1\@@endlink}%
\providecommand \@sanitize@url [0]{\catcode `\\12\catcode `\$12\catcode `\&12\catcode `\#12\catcode `\^12\catcode `\_12\catcode `\%12\relax}%
\providecommand \@@startlink[1]{}%
\providecommand \@@endlink[0]{}%
\providecommand \url  [0]{\begingroup\@sanitize@url \@url }%
\providecommand \@url [1]{\endgroup\@href {#1}{\urlprefix }}%
\providecommand \urlprefix  [0]{URL }%
\providecommand \Eprint [0]{\href }%
\providecommand \doibase [0]{https://doi.org/}%
\providecommand \selectlanguage [0]{\@gobble}%
\providecommand \bibinfo  [0]{\@secondoftwo}%
\providecommand \bibfield  [0]{\@secondoftwo}%
\providecommand \translation [1]{[#1]}%
\providecommand \BibitemOpen [0]{}%
\providecommand \bibitemStop [0]{}%
\providecommand \bibitemNoStop [0]{.\EOS\space}%
\providecommand \EOS [0]{\spacefactor3000\relax}%
\providecommand \BibitemShut  [1]{\csname bibitem#1\endcsname}%
\let\auto@bib@innerbib\@empty
\bibitem [{\citenamefont {Slussarenko}\ and\ \citenamefont {Pryde}(2019)}]{slussarenko2019photonic}%
  \BibitemOpen
  \bibfield  {author} {\bibinfo {author} {\bibfnamefont {S.}~\bibnamefont {Slussarenko}}\ and\ \bibinfo {author} {\bibfnamefont {G.~J.}\ \bibnamefont {Pryde}},\ }\bibfield  {title} {\bibinfo {title} {Photonic quantum information processing: A concise review},\ }\href@noop {} {\bibfield  {journal} {\bibinfo  {journal} {Applied physics reviews}\ }\textbf {\bibinfo {volume} {6}} (\bibinfo {year} {2019})}\BibitemShut {NoStop}%
\bibitem [{\citenamefont {Flamini}\ \emph {et~al.}(2018)\citenamefont {Flamini}, \citenamefont {Spagnolo},\ and\ \citenamefont {Sciarrino}}]{flamini2018photonic}%
  \BibitemOpen
  \bibfield  {author} {\bibinfo {author} {\bibfnamefont {F.}~\bibnamefont {Flamini}}, \bibinfo {author} {\bibfnamefont {N.}~\bibnamefont {Spagnolo}},\ and\ \bibinfo {author} {\bibfnamefont {F.}~\bibnamefont {Sciarrino}},\ }\bibfield  {title} {\bibinfo {title} {Photonic quantum information processing: a review},\ }\href@noop {} {\bibfield  {journal} {\bibinfo  {journal} {Reports on Progress in Physics}\ }\textbf {\bibinfo {volume} {82}},\ \bibinfo {pages} {016001} (\bibinfo {year} {2018})}\BibitemShut {NoStop}%
\bibitem [{\citenamefont {Couteau}\ \emph {et~al.}(2023)\citenamefont {Couteau}, \citenamefont {Barz}, \citenamefont {Durt}, \citenamefont {Gerrits}, \citenamefont {Huwer}, \citenamefont {Prevedel}, \citenamefont {Rarity}, \citenamefont {Shields},\ and\ \citenamefont {Weihs}}]{couteau2023applications}%
  \BibitemOpen
  \bibfield  {author} {\bibinfo {author} {\bibfnamefont {C.}~\bibnamefont {Couteau}}, \bibinfo {author} {\bibfnamefont {S.}~\bibnamefont {Barz}}, \bibinfo {author} {\bibfnamefont {T.}~\bibnamefont {Durt}}, \bibinfo {author} {\bibfnamefont {T.}~\bibnamefont {Gerrits}}, \bibinfo {author} {\bibfnamefont {J.}~\bibnamefont {Huwer}}, \bibinfo {author} {\bibfnamefont {R.}~\bibnamefont {Prevedel}}, \bibinfo {author} {\bibfnamefont {J.}~\bibnamefont {Rarity}}, \bibinfo {author} {\bibfnamefont {A.}~\bibnamefont {Shields}},\ and\ \bibinfo {author} {\bibfnamefont {G.}~\bibnamefont {Weihs}},\ }\bibfield  {title} {\bibinfo {title} {Applications of single photons to quantum communication and computing},\ }\href@noop {} {\bibfield  {journal} {\bibinfo  {journal} {Nature Reviews Physics}\ }\textbf {\bibinfo {volume} {5}},\ \bibinfo {pages} {326} (\bibinfo {year} {2023})}\BibitemShut {NoStop}%
\bibitem [{\citenamefont {Rohde}\ and\ \citenamefont {Ralph}(2006)}]{rohde2006error}%
  \BibitemOpen
  \bibfield  {author} {\bibinfo {author} {\bibfnamefont {P.~P.}\ \bibnamefont {Rohde}}\ and\ \bibinfo {author} {\bibfnamefont {T.~C.}\ \bibnamefont {Ralph}},\ }\bibfield  {title} {\bibinfo {title} {Error models for mode mismatch in linear optics quantum computing},\ }\href@noop {} {\bibfield  {journal} {\bibinfo  {journal} {Physical Review A—Atomic, Molecular, and Optical Physics}\ }\textbf {\bibinfo {volume} {73}},\ \bibinfo {pages} {062312} (\bibinfo {year} {2006})}\BibitemShut {NoStop}%
\bibitem [{\citenamefont {Renema}\ \emph {et~al.}(2018)\citenamefont {Renema}, \citenamefont {Menssen}, \citenamefont {Clements}, \citenamefont {Triginer}, \citenamefont {Kolthammer},\ and\ \citenamefont {Walmsley}}]{renema2018efficient}%
  \BibitemOpen
  \bibfield  {author} {\bibinfo {author} {\bibfnamefont {J.~J.}\ \bibnamefont {Renema}}, \bibinfo {author} {\bibfnamefont {A.}~\bibnamefont {Menssen}}, \bibinfo {author} {\bibfnamefont {W.~R.}\ \bibnamefont {Clements}}, \bibinfo {author} {\bibfnamefont {G.}~\bibnamefont {Triginer}}, \bibinfo {author} {\bibfnamefont {W.~S.}\ \bibnamefont {Kolthammer}},\ and\ \bibinfo {author} {\bibfnamefont {I.~A.}\ \bibnamefont {Walmsley}},\ }\bibfield  {title} {\bibinfo {title} {Efficient classical algorithm for boson sampling with partially distinguishable photons},\ }\href@noop {} {\bibfield  {journal} {\bibinfo  {journal} {Physical review letters}\ }\textbf {\bibinfo {volume} {120}},\ \bibinfo {pages} {220502} (\bibinfo {year} {2018})}\BibitemShut {NoStop}%
\bibitem [{\citenamefont {Saied}\ \emph {et~al.}(2024)\citenamefont {Saied}, \citenamefont {Marshall}, \citenamefont {Anand}, \citenamefont {Grabbe},\ and\ \citenamefont {Rieffel}}]{saied2024advancing}%
  \BibitemOpen
  \bibfield  {author} {\bibinfo {author} {\bibfnamefont {J.}~\bibnamefont {Saied}}, \bibinfo {author} {\bibfnamefont {J.}~\bibnamefont {Marshall}}, \bibinfo {author} {\bibfnamefont {N.}~\bibnamefont {Anand}}, \bibinfo {author} {\bibfnamefont {S.}~\bibnamefont {Grabbe}},\ and\ \bibinfo {author} {\bibfnamefont {E.~G.}\ \bibnamefont {Rieffel}},\ }\bibfield  {title} {\bibinfo {title} {Advancing quantum networking: some tools and protocols for ideal and noisy photonic systems},\ }in\ \href@noop {} {\emph {\bibinfo {booktitle} {Quantum Computing, Communication, and Simulation IV}}},\ Vol.\ \bibinfo {volume} {12911}\ (\bibinfo {organization} {SPIE},\ \bibinfo {year} {2024})\ pp.\ \bibinfo {pages} {37--64}\BibitemShut {NoStop}%
\bibitem [{\citenamefont {Shaw}\ \emph {et~al.}(2023)\citenamefont {Shaw}, \citenamefont {Jones}, \citenamefont {Yard},\ and\ \citenamefont {Laing}}]{shaw2023errors}%
  \BibitemOpen
  \bibfield  {author} {\bibinfo {author} {\bibfnamefont {R.~D.}\ \bibnamefont {Shaw}}, \bibinfo {author} {\bibfnamefont {A.~E.}\ \bibnamefont {Jones}}, \bibinfo {author} {\bibfnamefont {P.}~\bibnamefont {Yard}},\ and\ \bibinfo {author} {\bibfnamefont {A.}~\bibnamefont {Laing}},\ }\bibfield  {title} {\bibinfo {title} {Errors in heralded circuits for linear optical entanglement generation},\ }\href@noop {} {\bibfield  {journal} {\bibinfo  {journal} {arXiv preprint arXiv:2305.08452}\ } (\bibinfo {year} {2023})}\BibitemShut {NoStop}%
\bibitem [{\citenamefont {Glauber}(1963)}]{glauber1963quantum}%
  \BibitemOpen
  \bibfield  {author} {\bibinfo {author} {\bibfnamefont {R.~J.}\ \bibnamefont {Glauber}},\ }\bibfield  {title} {\bibinfo {title} {The quantum theory of optical coherence},\ }\href@noop {} {\bibfield  {journal} {\bibinfo  {journal} {Physical Review}\ }\textbf {\bibinfo {volume} {130}},\ \bibinfo {pages} {2529} (\bibinfo {year} {1963})}\BibitemShut {NoStop}%
\bibitem [{\citenamefont {Gr{\"u}nwald}(2019)}]{grunwald2019effective}%
  \BibitemOpen
  \bibfield  {author} {\bibinfo {author} {\bibfnamefont {P.}~\bibnamefont {Gr{\"u}nwald}},\ }\bibfield  {title} {\bibinfo {title} {Effective second-order correlation function and single-photon detection},\ }\href@noop {} {\bibfield  {journal} {\bibinfo  {journal} {New Journal of Physics}\ }\textbf {\bibinfo {volume} {21}},\ \bibinfo {pages} {093003} (\bibinfo {year} {2019})}\BibitemShut {NoStop}%
\bibitem [{\citenamefont {Hong}\ \emph {et~al.}(1987)\citenamefont {Hong}, \citenamefont {Ou},\ and\ \citenamefont {Mandel}}]{hong1987measurement}%
  \BibitemOpen
  \bibfield  {author} {\bibinfo {author} {\bibfnamefont {C.-K.}\ \bibnamefont {Hong}}, \bibinfo {author} {\bibfnamefont {Z.-Y.}\ \bibnamefont {Ou}},\ and\ \bibinfo {author} {\bibfnamefont {L.}~\bibnamefont {Mandel}},\ }\bibfield  {title} {\bibinfo {title} {Measurement of subpicosecond time intervals between two photons by interference},\ }\href {https://doi.org/https://doi.org/10.1103/PhysRevLett.59.2044} {\bibfield  {journal} {\bibinfo  {journal} {Physical Review Letters}\ }\textbf {\bibinfo {volume} {59}},\ \bibinfo {pages} {2044} (\bibinfo {year} {1987})}\BibitemShut {NoStop}%
\bibitem [{\citenamefont {Marshall}(2022)}]{marshall2022distillation}%
  \BibitemOpen
  \bibfield  {author} {\bibinfo {author} {\bibfnamefont {J.}~\bibnamefont {Marshall}},\ }\bibfield  {title} {\bibinfo {title} {Distillation of indistinguishable photons},\ }\href@noop {} {\bibfield  {journal} {\bibinfo  {journal} {Physical Review Letters}\ }\textbf {\bibinfo {volume} {129}},\ \bibinfo {pages} {213601} (\bibinfo {year} {2022})}\BibitemShut {NoStop}%
\bibitem [{\citenamefont {Trivedi}\ \emph {et~al.}(2020)\citenamefont {Trivedi}, \citenamefont {Fischer}, \citenamefont {Vu{\v{c}}kovi{\'c}},\ and\ \citenamefont {M{\"u}ller}}]{trivedi2020generation}%
  \BibitemOpen
  \bibfield  {author} {\bibinfo {author} {\bibfnamefont {R.}~\bibnamefont {Trivedi}}, \bibinfo {author} {\bibfnamefont {K.~A.}\ \bibnamefont {Fischer}}, \bibinfo {author} {\bibfnamefont {J.}~\bibnamefont {Vu{\v{c}}kovi{\'c}}},\ and\ \bibinfo {author} {\bibfnamefont {K.}~\bibnamefont {M{\"u}ller}},\ }\bibfield  {title} {\bibinfo {title} {Generation of non-classical light using semiconductor quantum dots},\ }\href@noop {} {\bibfield  {journal} {\bibinfo  {journal} {Advanced Quantum Technologies}\ }\textbf {\bibinfo {volume} {3}},\ \bibinfo {pages} {1900007} (\bibinfo {year} {2020})}\BibitemShut {NoStop}%
\bibitem [{\citenamefont {Somaschi}\ \emph {et~al.}(2016)\citenamefont {Somaschi}, \citenamefont {Giesz}, \citenamefont {De~Santis}, \citenamefont {Loredo}, \citenamefont {Almeida}, \citenamefont {Hornecker}, \citenamefont {Portalupi}, \citenamefont {Grange}, \citenamefont {Anton}, \citenamefont {Demory} \emph {et~al.}}]{somaschi2016near}%
  \BibitemOpen
  \bibfield  {author} {\bibinfo {author} {\bibfnamefont {N.}~\bibnamefont {Somaschi}}, \bibinfo {author} {\bibfnamefont {V.}~\bibnamefont {Giesz}}, \bibinfo {author} {\bibfnamefont {L.}~\bibnamefont {De~Santis}}, \bibinfo {author} {\bibfnamefont {J.}~\bibnamefont {Loredo}}, \bibinfo {author} {\bibfnamefont {M.~P.}\ \bibnamefont {Almeida}}, \bibinfo {author} {\bibfnamefont {G.}~\bibnamefont {Hornecker}}, \bibinfo {author} {\bibfnamefont {S.~L.}\ \bibnamefont {Portalupi}}, \bibinfo {author} {\bibfnamefont {T.}~\bibnamefont {Grange}}, \bibinfo {author} {\bibfnamefont {C.}~\bibnamefont {Anton}}, \bibinfo {author} {\bibfnamefont {J.}~\bibnamefont {Demory}}, \emph {et~al.},\ }\bibfield  {title} {\bibinfo {title} {Near-optimal single-photon sources in the solid state},\ }\href@noop {} {\bibfield  {journal} {\bibinfo  {journal} {Nature Photonics}\ }\textbf {\bibinfo {volume} {10}},\ \bibinfo {pages} {340} (\bibinfo {year} {2016})}\BibitemShut {NoStop}%
\bibitem [{\citenamefont {Wang}\ \emph {et~al.}(2019)\citenamefont {Wang}, \citenamefont {He}, \citenamefont {Chung}, \citenamefont {Hu}, \citenamefont {Yu}, \citenamefont {Chen}, \citenamefont {Ding}, \citenamefont {Chen}, \citenamefont {Qin}, \citenamefont {Yang} \emph {et~al.}}]{wang2019towards}%
  \BibitemOpen
  \bibfield  {author} {\bibinfo {author} {\bibfnamefont {H.}~\bibnamefont {Wang}}, \bibinfo {author} {\bibfnamefont {Y.-M.}\ \bibnamefont {He}}, \bibinfo {author} {\bibfnamefont {T.-H.}\ \bibnamefont {Chung}}, \bibinfo {author} {\bibfnamefont {H.}~\bibnamefont {Hu}}, \bibinfo {author} {\bibfnamefont {Y.}~\bibnamefont {Yu}}, \bibinfo {author} {\bibfnamefont {S.}~\bibnamefont {Chen}}, \bibinfo {author} {\bibfnamefont {X.}~\bibnamefont {Ding}}, \bibinfo {author} {\bibfnamefont {M.-C.}\ \bibnamefont {Chen}}, \bibinfo {author} {\bibfnamefont {J.}~\bibnamefont {Qin}}, \bibinfo {author} {\bibfnamefont {X.}~\bibnamefont {Yang}}, \emph {et~al.},\ }\bibfield  {title} {\bibinfo {title} {Towards optimal single-photon sources from polarized microcavities},\ }\href@noop {} {\bibfield  {journal} {\bibinfo  {journal} {Nature Photonics}\ }\textbf {\bibinfo {volume} {13}},\ \bibinfo {pages} {770} (\bibinfo {year} {2019})}\BibitemShut {NoStop}%
\bibitem [{\citenamefont {Grange}\ \emph {et~al.}(2017)\citenamefont {Grange}, \citenamefont {Somaschi}, \citenamefont {Ant{\'o}n}, \citenamefont {De~Santis}, \citenamefont {Coppola}, \citenamefont {Giesz}, \citenamefont {Lema{\^\i}tre}, \citenamefont {Sagnes}, \citenamefont {Auff{\`e}ves},\ and\ \citenamefont {Senellart}}]{grange2017reducing}%
  \BibitemOpen
  \bibfield  {author} {\bibinfo {author} {\bibfnamefont {T.}~\bibnamefont {Grange}}, \bibinfo {author} {\bibfnamefont {N.}~\bibnamefont {Somaschi}}, \bibinfo {author} {\bibfnamefont {C.}~\bibnamefont {Ant{\'o}n}}, \bibinfo {author} {\bibfnamefont {L.}~\bibnamefont {De~Santis}}, \bibinfo {author} {\bibfnamefont {G.}~\bibnamefont {Coppola}}, \bibinfo {author} {\bibfnamefont {V.}~\bibnamefont {Giesz}}, \bibinfo {author} {\bibfnamefont {A.}~\bibnamefont {Lema{\^\i}tre}}, \bibinfo {author} {\bibfnamefont {I.}~\bibnamefont {Sagnes}}, \bibinfo {author} {\bibfnamefont {A.}~\bibnamefont {Auff{\`e}ves}},\ and\ \bibinfo {author} {\bibfnamefont {P.}~\bibnamefont {Senellart}},\ }\bibfield  {title} {\bibinfo {title} {Reducing phonon-induced decoherence in solid-state single-photon sources with cavity quantum electrodynamics},\ }\href@noop {} {\bibfield  {journal} {\bibinfo  {journal} {Physical review letters}\ }\textbf {\bibinfo {volume} {118}},\ \bibinfo {pages} {253602} (\bibinfo {year} {2017})}\BibitemShut {NoStop}%
\bibitem [{\citenamefont {Kir{\v{s}}ansk{\.e}}\ \emph {et~al.}(2017)\citenamefont {Kir{\v{s}}ansk{\.e}}, \citenamefont {Thyrrestrup}, \citenamefont {Daveau}, \citenamefont {Dree{\ss}en}, \citenamefont {Pregnolato}, \citenamefont {Midolo}, \citenamefont {Tighineanu}, \citenamefont {Javadi}, \citenamefont {Stobbe}, \citenamefont {Schott} \emph {et~al.}}]{kirvsanske2017indistinguishable}%
  \BibitemOpen
  \bibfield  {author} {\bibinfo {author} {\bibfnamefont {G.}~\bibnamefont {Kir{\v{s}}ansk{\.e}}}, \bibinfo {author} {\bibfnamefont {H.}~\bibnamefont {Thyrrestrup}}, \bibinfo {author} {\bibfnamefont {R.~S.}\ \bibnamefont {Daveau}}, \bibinfo {author} {\bibfnamefont {C.~L.}\ \bibnamefont {Dree{\ss}en}}, \bibinfo {author} {\bibfnamefont {T.}~\bibnamefont {Pregnolato}}, \bibinfo {author} {\bibfnamefont {L.}~\bibnamefont {Midolo}}, \bibinfo {author} {\bibfnamefont {P.}~\bibnamefont {Tighineanu}}, \bibinfo {author} {\bibfnamefont {A.}~\bibnamefont {Javadi}}, \bibinfo {author} {\bibfnamefont {S.}~\bibnamefont {Stobbe}}, \bibinfo {author} {\bibfnamefont {R.}~\bibnamefont {Schott}}, \emph {et~al.},\ }\bibfield  {title} {\bibinfo {title} {Indistinguishable and efficient single photons from a quantum dot in a planar nanobeam waveguide},\ }\href@noop {} {\bibfield  {journal} {\bibinfo  {journal} {Physical Review B}\ }\textbf {\bibinfo {volume} {96}},\ \bibinfo {pages} {165306} (\bibinfo {year} {2017})}\BibitemShut
  {NoStop}%
\bibitem [{\citenamefont {Tomm}\ \emph {et~al.}(2021)\citenamefont {Tomm}, \citenamefont {Javadi}, \citenamefont {Antoniadis}, \citenamefont {Najer}, \citenamefont {L{\"o}bl}, \citenamefont {Korsch}, \citenamefont {Schott}, \citenamefont {Valentin}, \citenamefont {Wieck}, \citenamefont {Ludwig} \emph {et~al.}}]{tomm2021bright}%
  \BibitemOpen
  \bibfield  {author} {\bibinfo {author} {\bibfnamefont {N.}~\bibnamefont {Tomm}}, \bibinfo {author} {\bibfnamefont {A.}~\bibnamefont {Javadi}}, \bibinfo {author} {\bibfnamefont {N.~O.}\ \bibnamefont {Antoniadis}}, \bibinfo {author} {\bibfnamefont {D.}~\bibnamefont {Najer}}, \bibinfo {author} {\bibfnamefont {M.~C.}\ \bibnamefont {L{\"o}bl}}, \bibinfo {author} {\bibfnamefont {A.~R.}\ \bibnamefont {Korsch}}, \bibinfo {author} {\bibfnamefont {R.}~\bibnamefont {Schott}}, \bibinfo {author} {\bibfnamefont {S.~R.}\ \bibnamefont {Valentin}}, \bibinfo {author} {\bibfnamefont {A.~D.}\ \bibnamefont {Wieck}}, \bibinfo {author} {\bibfnamefont {A.}~\bibnamefont {Ludwig}}, \emph {et~al.},\ }\bibfield  {title} {\bibinfo {title} {A bright and fast source of coherent single photons},\ }\href {https://doi.org/https://doi.org/10.1038/s41565-020-00831-x} {\bibfield  {journal} {\bibinfo  {journal} {Nature Nanotechnology}\ }\textbf {\bibinfo {volume} {16}},\ \bibinfo {pages} {399} (\bibinfo {year} {2021})}\BibitemShut {NoStop}%
\bibitem [{\citenamefont {Ollivier}\ \emph {et~al.}(2021)\citenamefont {Ollivier}, \citenamefont {Thomas}, \citenamefont {Wein}, \citenamefont {de~Buy~Wenniger}, \citenamefont {Coste}, \citenamefont {Loredo}, \citenamefont {Somaschi}, \citenamefont {Harouri}, \citenamefont {Lemaitre}, \citenamefont {Sagnes} \emph {et~al.}}]{ollivier2021hong}%
  \BibitemOpen
  \bibfield  {author} {\bibinfo {author} {\bibfnamefont {H.}~\bibnamefont {Ollivier}}, \bibinfo {author} {\bibfnamefont {S.}~\bibnamefont {Thomas}}, \bibinfo {author} {\bibfnamefont {S.}~\bibnamefont {Wein}}, \bibinfo {author} {\bibfnamefont {I.~M.}\ \bibnamefont {de~Buy~Wenniger}}, \bibinfo {author} {\bibfnamefont {N.}~\bibnamefont {Coste}}, \bibinfo {author} {\bibfnamefont {J.}~\bibnamefont {Loredo}}, \bibinfo {author} {\bibfnamefont {N.}~\bibnamefont {Somaschi}}, \bibinfo {author} {\bibfnamefont {A.}~\bibnamefont {Harouri}}, \bibinfo {author} {\bibfnamefont {A.}~\bibnamefont {Lemaitre}}, \bibinfo {author} {\bibfnamefont {I.}~\bibnamefont {Sagnes}}, \emph {et~al.},\ }\bibfield  {title} {\bibinfo {title} {Hong-ou-mandel interference with imperfect single photon sources},\ }\href {https://doi.org/10.1103/PhysRevLett.126.063602} {\bibfield  {journal} {\bibinfo  {journal} {Physical Review Letters}\ }\textbf {\bibinfo {volume} {126}},\ \bibinfo {pages} {063602} (\bibinfo {year} {2021})}\BibitemShut {NoStop}%
\bibitem [{\citenamefont {Gonz{\'a}lez-Ruiz}\ \emph {et~al.}(2025)\citenamefont {Gonz{\'a}lez-Ruiz}, \citenamefont {Bjerlin}, \citenamefont {Sandberg},\ and\ \citenamefont {S{\o}rensen}}]{gonzalez2025two}%
  \BibitemOpen
  \bibfield  {author} {\bibinfo {author} {\bibfnamefont {E.~M.}\ \bibnamefont {Gonz{\'a}lez-Ruiz}}, \bibinfo {author} {\bibfnamefont {J.}~\bibnamefont {Bjerlin}}, \bibinfo {author} {\bibfnamefont {O.~A.~D.}\ \bibnamefont {Sandberg}},\ and\ \bibinfo {author} {\bibfnamefont {A.~S.}\ \bibnamefont {S{\o}rensen}},\ }\bibfield  {title} {\bibinfo {title} {Two-photon correlations and hong-ou-mandel visibility from an imperfect single-photon source},\ }\href@noop {} {\bibfield  {journal} {\bibinfo  {journal} {Physical Review Applied}\ }\textbf {\bibinfo {volume} {23}},\ \bibinfo {pages} {054063} (\bibinfo {year} {2025})}\BibitemShut {NoStop}%
\bibitem [{\citenamefont {Ding}\ \emph {et~al.}(2025)\citenamefont {Ding}, \citenamefont {Guo}, \citenamefont {Xu}, \citenamefont {Liu}, \citenamefont {Zou}, \citenamefont {Zhao}, \citenamefont {Ge}, \citenamefont {Zhang}, \citenamefont {Liu}, \citenamefont {Wang} \emph {et~al.}}]{ding2025high}%
  \BibitemOpen
  \bibfield  {author} {\bibinfo {author} {\bibfnamefont {X.}~\bibnamefont {Ding}}, \bibinfo {author} {\bibfnamefont {Y.-P.}\ \bibnamefont {Guo}}, \bibinfo {author} {\bibfnamefont {M.-C.}\ \bibnamefont {Xu}}, \bibinfo {author} {\bibfnamefont {R.-Z.}\ \bibnamefont {Liu}}, \bibinfo {author} {\bibfnamefont {G.-Y.}\ \bibnamefont {Zou}}, \bibinfo {author} {\bibfnamefont {J.-Y.}\ \bibnamefont {Zhao}}, \bibinfo {author} {\bibfnamefont {Z.-X.}\ \bibnamefont {Ge}}, \bibinfo {author} {\bibfnamefont {Q.-H.}\ \bibnamefont {Zhang}}, \bibinfo {author} {\bibfnamefont {H.-L.}\ \bibnamefont {Liu}}, \bibinfo {author} {\bibfnamefont {L.-J.}\ \bibnamefont {Wang}}, \emph {et~al.},\ }\bibfield  {title} {\bibinfo {title} {High-efficiency single-photon source above the loss-tolerant threshold for efficient linear optical quantum computing},\ }\href@noop {} {\bibfield  {journal} {\bibinfo  {journal} {Nature Photonics}\ ,\ \bibinfo {pages} {1}} (\bibinfo {year} {2025})}\BibitemShut {NoStop}%
\bibitem [{\citenamefont {Michler}\ \emph {et~al.}(2000)\citenamefont {Michler}, \citenamefont {Kiraz}, \citenamefont {Becher}, \citenamefont {Schoenfeld}, \citenamefont {Petroff}, \citenamefont {Zhang}, \citenamefont {Hu},\ and\ \citenamefont {Imamoglu}}]{michler2000quantum}%
  \BibitemOpen
  \bibfield  {author} {\bibinfo {author} {\bibfnamefont {P.}~\bibnamefont {Michler}}, \bibinfo {author} {\bibfnamefont {A.}~\bibnamefont {Kiraz}}, \bibinfo {author} {\bibfnamefont {C.}~\bibnamefont {Becher}}, \bibinfo {author} {\bibfnamefont {W.}~\bibnamefont {Schoenfeld}}, \bibinfo {author} {\bibfnamefont {P.}~\bibnamefont {Petroff}}, \bibinfo {author} {\bibfnamefont {L.}~\bibnamefont {Zhang}}, \bibinfo {author} {\bibfnamefont {E.}~\bibnamefont {Hu}},\ and\ \bibinfo {author} {\bibfnamefont {A.}~\bibnamefont {Imamoglu}},\ }\bibfield  {title} {\bibinfo {title} {A quantum dot single-photon turnstile device},\ }\href@noop {} {\bibfield  {journal} {\bibinfo  {journal} {science}\ }\textbf {\bibinfo {volume} {290}},\ \bibinfo {pages} {2282} (\bibinfo {year} {2000})}\BibitemShut {NoStop}%
\bibitem [{\citenamefont {Ornelas-Huerta}\ \emph {et~al.}(2020)\citenamefont {Ornelas-Huerta}, \citenamefont {Craddock}, \citenamefont {Goldschmidt}, \citenamefont {Hachtel}, \citenamefont {Wang}, \citenamefont {Bienias}, \citenamefont {Gorshkov}, \citenamefont {Rolston},\ and\ \citenamefont {Porto}}]{ornelas2020demand}%
  \BibitemOpen
  \bibfield  {author} {\bibinfo {author} {\bibfnamefont {D.~P.}\ \bibnamefont {Ornelas-Huerta}}, \bibinfo {author} {\bibfnamefont {A.~N.}\ \bibnamefont {Craddock}}, \bibinfo {author} {\bibfnamefont {E.~A.}\ \bibnamefont {Goldschmidt}}, \bibinfo {author} {\bibfnamefont {A.~J.}\ \bibnamefont {Hachtel}}, \bibinfo {author} {\bibfnamefont {Y.}~\bibnamefont {Wang}}, \bibinfo {author} {\bibfnamefont {P.}~\bibnamefont {Bienias}}, \bibinfo {author} {\bibfnamefont {A.~V.}\ \bibnamefont {Gorshkov}}, \bibinfo {author} {\bibfnamefont {S.~L.}\ \bibnamefont {Rolston}},\ and\ \bibinfo {author} {\bibfnamefont {J.~V.}\ \bibnamefont {Porto}},\ }\bibfield  {title} {\bibinfo {title} {On-demand indistinguishable single photons from an efficient and pure source based on a rydberg ensemble},\ }\href@noop {} {\bibfield  {journal} {\bibinfo  {journal} {Optica}\ }\textbf {\bibinfo {volume} {7}},\ \bibinfo {pages} {813} (\bibinfo {year} {2020})}\BibitemShut {NoStop}%
\bibitem [{\citenamefont {Fournier}\ \emph {et~al.}(2023)\citenamefont {Fournier}, \citenamefont {Roux}, \citenamefont {Watanabe}, \citenamefont {Taniguchi}, \citenamefont {Buil}, \citenamefont {Barjon}, \citenamefont {Hermier},\ and\ \citenamefont {Delteil}}]{fournier2023two}%
  \BibitemOpen
  \bibfield  {author} {\bibinfo {author} {\bibfnamefont {C.}~\bibnamefont {Fournier}}, \bibinfo {author} {\bibfnamefont {S.}~\bibnamefont {Roux}}, \bibinfo {author} {\bibfnamefont {K.}~\bibnamefont {Watanabe}}, \bibinfo {author} {\bibfnamefont {T.}~\bibnamefont {Taniguchi}}, \bibinfo {author} {\bibfnamefont {S.}~\bibnamefont {Buil}}, \bibinfo {author} {\bibfnamefont {J.}~\bibnamefont {Barjon}}, \bibinfo {author} {\bibfnamefont {J.-P.}\ \bibnamefont {Hermier}},\ and\ \bibinfo {author} {\bibfnamefont {A.}~\bibnamefont {Delteil}},\ }\bibfield  {title} {\bibinfo {title} {Two-photon interference from a quantum emitter in hexagonal boron nitride},\ }\href@noop {} {\bibfield  {journal} {\bibinfo  {journal} {Physical Review Applied}\ }\textbf {\bibinfo {volume} {19}},\ \bibinfo {pages} {L041003} (\bibinfo {year} {2023})}\BibitemShut {NoStop}%
\bibitem [{\citenamefont {Bernien}\ \emph {et~al.}(2013)\citenamefont {Bernien}, \citenamefont {Hensen}, \citenamefont {Pfaff}, \citenamefont {Koolstra}, \citenamefont {Blok}, \citenamefont {Robledo}, \citenamefont {Taminiau}, \citenamefont {Markham}, \citenamefont {Twitchen}, \citenamefont {Childress} \emph {et~al.}}]{bernien2013heralded}%
  \BibitemOpen
  \bibfield  {author} {\bibinfo {author} {\bibfnamefont {H.}~\bibnamefont {Bernien}}, \bibinfo {author} {\bibfnamefont {B.}~\bibnamefont {Hensen}}, \bibinfo {author} {\bibfnamefont {W.}~\bibnamefont {Pfaff}}, \bibinfo {author} {\bibfnamefont {G.}~\bibnamefont {Koolstra}}, \bibinfo {author} {\bibfnamefont {M.~S.}\ \bibnamefont {Blok}}, \bibinfo {author} {\bibfnamefont {L.}~\bibnamefont {Robledo}}, \bibinfo {author} {\bibfnamefont {T.~H.}\ \bibnamefont {Taminiau}}, \bibinfo {author} {\bibfnamefont {M.}~\bibnamefont {Markham}}, \bibinfo {author} {\bibfnamefont {D.~J.}\ \bibnamefont {Twitchen}}, \bibinfo {author} {\bibfnamefont {L.}~\bibnamefont {Childress}}, \emph {et~al.},\ }\bibfield  {title} {\bibinfo {title} {Heralded entanglement between solid-state qubits separated by three metres},\ }\href@noop {} {\bibfield  {journal} {\bibinfo  {journal} {Nature}\ }\textbf {\bibinfo {volume} {497}},\ \bibinfo {pages} {86} (\bibinfo {year} {2013})}\BibitemShut {NoStop}%
\bibitem [{\citenamefont {Pont}\ \emph {et~al.}(2024)\citenamefont {Pont}, \citenamefont {Corrielli}, \citenamefont {Fyrillas}, \citenamefont {Agresti}, \citenamefont {Carvacho}, \citenamefont {Maring}, \citenamefont {Emeriau}, \citenamefont {Ceccarelli}, \citenamefont {Albiero}, \citenamefont {Dias~Ferreira} \emph {et~al.}}]{pont2024high}%
  \BibitemOpen
  \bibfield  {author} {\bibinfo {author} {\bibfnamefont {M.}~\bibnamefont {Pont}}, \bibinfo {author} {\bibfnamefont {G.}~\bibnamefont {Corrielli}}, \bibinfo {author} {\bibfnamefont {A.}~\bibnamefont {Fyrillas}}, \bibinfo {author} {\bibfnamefont {I.}~\bibnamefont {Agresti}}, \bibinfo {author} {\bibfnamefont {G.}~\bibnamefont {Carvacho}}, \bibinfo {author} {\bibfnamefont {N.}~\bibnamefont {Maring}}, \bibinfo {author} {\bibfnamefont {P.-E.}\ \bibnamefont {Emeriau}}, \bibinfo {author} {\bibfnamefont {F.}~\bibnamefont {Ceccarelli}}, \bibinfo {author} {\bibfnamefont {R.}~\bibnamefont {Albiero}}, \bibinfo {author} {\bibfnamefont {P.~H.}\ \bibnamefont {Dias~Ferreira}}, \emph {et~al.},\ }\bibfield  {title} {\bibinfo {title} {High-fidelity four-photon ghz states on chip},\ }\href@noop {} {\bibfield  {journal} {\bibinfo  {journal} {npj Quantum Information}\ }\textbf {\bibinfo {volume} {10}},\ \bibinfo {pages} {50} (\bibinfo {year} {2024})}\BibitemShut {NoStop}%
\bibitem [{\citenamefont {Oszmaniec}\ and\ \citenamefont {Brod}(2018)}]{oszmaniec2018classical}%
  \BibitemOpen
  \bibfield  {author} {\bibinfo {author} {\bibfnamefont {M.}~\bibnamefont {Oszmaniec}}\ and\ \bibinfo {author} {\bibfnamefont {D.~J.}\ \bibnamefont {Brod}},\ }\bibfield  {title} {\bibinfo {title} {Classical simulation of photonic linear optics with lost particles},\ }\href@noop {} {\bibfield  {journal} {\bibinfo  {journal} {New Journal of Physics}\ }\textbf {\bibinfo {volume} {20}},\ \bibinfo {pages} {092002} (\bibinfo {year} {2018})}\BibitemShut {NoStop}%
\bibitem [{\citenamefont {Annoni}\ and\ \citenamefont {Wein}(2025)}]{annoni2025incoherent}%
  \BibitemOpen
  \bibfield  {author} {\bibinfo {author} {\bibfnamefont {E.}~\bibnamefont {Annoni}}\ and\ \bibinfo {author} {\bibfnamefont {S.~C.}\ \bibnamefont {Wein}},\ }\bibfield  {title} {\bibinfo {title} {Incoherent behavior of partially distinguishable photons},\ }\href@noop {} {\bibfield  {journal} {\bibinfo  {journal} {arXiv preprint arXiv:2502.05047}\ } (\bibinfo {year} {2025})}\BibitemShut {NoStop}%
\bibitem [{\citenamefont {Menssen}\ \emph {et~al.}(2017)\citenamefont {Menssen}, \citenamefont {Jones}, \citenamefont {Metcalf}, \citenamefont {Tichy}, \citenamefont {Barz}, \citenamefont {Kolthammer},\ and\ \citenamefont {Walmsley}}]{menssen2017distinguishability}%
  \BibitemOpen
  \bibfield  {author} {\bibinfo {author} {\bibfnamefont {A.~J.}\ \bibnamefont {Menssen}}, \bibinfo {author} {\bibfnamefont {A.~E.}\ \bibnamefont {Jones}}, \bibinfo {author} {\bibfnamefont {B.~J.}\ \bibnamefont {Metcalf}}, \bibinfo {author} {\bibfnamefont {M.~C.}\ \bibnamefont {Tichy}}, \bibinfo {author} {\bibfnamefont {S.}~\bibnamefont {Barz}}, \bibinfo {author} {\bibfnamefont {W.~S.}\ \bibnamefont {Kolthammer}},\ and\ \bibinfo {author} {\bibfnamefont {I.~A.}\ \bibnamefont {Walmsley}},\ }\bibfield  {title} {\bibinfo {title} {Distinguishability and many-particle interference},\ }\href@noop {} {\bibfield  {journal} {\bibinfo  {journal} {Physical review letters}\ }\textbf {\bibinfo {volume} {118}},\ \bibinfo {pages} {153603} (\bibinfo {year} {2017})}\BibitemShut {NoStop}%
\bibitem [{\citenamefont {Seron}\ \emph {et~al.}(2023)\citenamefont {Seron}, \citenamefont {Novo},\ and\ \citenamefont {Cerf}}]{seron2023boson}%
  \BibitemOpen
  \bibfield  {author} {\bibinfo {author} {\bibfnamefont {B.}~\bibnamefont {Seron}}, \bibinfo {author} {\bibfnamefont {L.}~\bibnamefont {Novo}},\ and\ \bibinfo {author} {\bibfnamefont {N.~J.}\ \bibnamefont {Cerf}},\ }\bibfield  {title} {\bibinfo {title} {Boson bunching is not maximized by indistinguishable particles},\ }\href@noop {} {\bibfield  {journal} {\bibinfo  {journal} {Nature Photonics}\ }\textbf {\bibinfo {volume} {17}},\ \bibinfo {pages} {702} (\bibinfo {year} {2023})}\BibitemShut {NoStop}%
\bibitem [{\citenamefont {Sparrow}(2017)}]{sparrow2017quantum}%
  \BibitemOpen
  \bibfield  {author} {\bibinfo {author} {\bibfnamefont {C.}~\bibnamefont {Sparrow}},\ }\emph {\bibinfo {title} {Quantum interference in universal linear optical devices for quantum computation and simulation}},\ \href {https://doi.org/https://doi.org/10.25560/67638} {Ph.D. thesis} (\bibinfo {year} {2017})\BibitemShut {NoStop}%
\bibitem [{\citenamefont {Tsujimoto}\ \emph {et~al.}(2021)\citenamefont {Tsujimoto}, \citenamefont {Wakui}, \citenamefont {Fujiwara}, \citenamefont {Sasaki},\ and\ \citenamefont {Takeoka}}]{tsujimoto2021ultra}%
  \BibitemOpen
  \bibfield  {author} {\bibinfo {author} {\bibfnamefont {Y.}~\bibnamefont {Tsujimoto}}, \bibinfo {author} {\bibfnamefont {K.}~\bibnamefont {Wakui}}, \bibinfo {author} {\bibfnamefont {M.}~\bibnamefont {Fujiwara}}, \bibinfo {author} {\bibfnamefont {M.}~\bibnamefont {Sasaki}},\ and\ \bibinfo {author} {\bibfnamefont {M.}~\bibnamefont {Takeoka}},\ }\bibfield  {title} {\bibinfo {title} {Ultra-fast hong-ou-mandel interferometry via temporal filtering},\ }\href@noop {} {\bibfield  {journal} {\bibinfo  {journal} {Optics Express}\ }\textbf {\bibinfo {volume} {29}},\ \bibinfo {pages} {37150} (\bibinfo {year} {2021})}\BibitemShut {NoStop}%
\bibitem [{\citenamefont {Tsujimoto}\ \emph {et~al.}(2023)\citenamefont {Tsujimoto}, \citenamefont {Ikuta}, \citenamefont {Wakui}, \citenamefont {Kobayashi},\ and\ \citenamefont {Fujiwara}}]{tsujimoto2023quantum}%
  \BibitemOpen
  \bibfield  {author} {\bibinfo {author} {\bibfnamefont {Y.}~\bibnamefont {Tsujimoto}}, \bibinfo {author} {\bibfnamefont {R.}~\bibnamefont {Ikuta}}, \bibinfo {author} {\bibfnamefont {K.}~\bibnamefont {Wakui}}, \bibinfo {author} {\bibfnamefont {T.}~\bibnamefont {Kobayashi}},\ and\ \bibinfo {author} {\bibfnamefont {M.}~\bibnamefont {Fujiwara}},\ }\bibfield  {title} {\bibinfo {title} {Quantum state tomography of qudits via hong-ou-mandel interference},\ }\href@noop {} {\bibfield  {journal} {\bibinfo  {journal} {Physical Review Applied}\ }\textbf {\bibinfo {volume} {19}},\ \bibinfo {pages} {014008} (\bibinfo {year} {2023})}\BibitemShut {NoStop}%
\bibitem [{\citenamefont {Renema}\ \emph {et~al.}(2021)\citenamefont {Renema}, \citenamefont {Wang}, \citenamefont {Qin}, \citenamefont {You}, \citenamefont {Lu},\ and\ \citenamefont {Pan}}]{renema2021sample}%
  \BibitemOpen
  \bibfield  {author} {\bibinfo {author} {\bibfnamefont {J.~J.}\ \bibnamefont {Renema}}, \bibinfo {author} {\bibfnamefont {H.}~\bibnamefont {Wang}}, \bibinfo {author} {\bibfnamefont {J.}~\bibnamefont {Qin}}, \bibinfo {author} {\bibfnamefont {X.}~\bibnamefont {You}}, \bibinfo {author} {\bibfnamefont {C.}~\bibnamefont {Lu}},\ and\ \bibinfo {author} {\bibfnamefont {J.}~\bibnamefont {Pan}},\ }\bibfield  {title} {\bibinfo {title} {Sample-efficient benchmarking of multiphoton interference on a boson sampler in the sparse regime},\ }\href@noop {} {\bibfield  {journal} {\bibinfo  {journal} {Physical Review A}\ }\textbf {\bibinfo {volume} {103}},\ \bibinfo {pages} {023722} (\bibinfo {year} {2021})}\BibitemShut {NoStop}%
\bibitem [{\citenamefont {Moylett}\ \emph {et~al.}(2019)\citenamefont {Moylett}, \citenamefont {Garc{\'\i}a-Patr{\'o}n}, \citenamefont {Renema},\ and\ \citenamefont {Turner}}]{moylett2019classically}%
  \BibitemOpen
  \bibfield  {author} {\bibinfo {author} {\bibfnamefont {A.~E.}\ \bibnamefont {Moylett}}, \bibinfo {author} {\bibfnamefont {R.}~\bibnamefont {Garc{\'\i}a-Patr{\'o}n}}, \bibinfo {author} {\bibfnamefont {J.~J.}\ \bibnamefont {Renema}},\ and\ \bibinfo {author} {\bibfnamefont {P.~S.}\ \bibnamefont {Turner}},\ }\bibfield  {title} {\bibinfo {title} {Classically simulating near-term partially-distinguishable and lossy boson sampling},\ }\href {https://doi.org/10.1088/2058-9565/ab5555} {\bibfield  {journal} {\bibinfo  {journal} {Quantum Science and Technology}\ }\textbf {\bibinfo {volume} {5}},\ \bibinfo {pages} {015001} (\bibinfo {year} {2019})}\BibitemShut {NoStop}%
\bibitem [{\citenamefont {Somhorst}\ \emph {et~al.}(2025)\citenamefont {Somhorst}, \citenamefont {Sau{\"e}r}, \citenamefont {van~den Hoven},\ and\ \citenamefont {Renema}}]{somhorst2025photon}%
  \BibitemOpen
  \bibfield  {author} {\bibinfo {author} {\bibfnamefont {F.}~\bibnamefont {Somhorst}}, \bibinfo {author} {\bibfnamefont {B.~K.}\ \bibnamefont {Sau{\"e}r}}, \bibinfo {author} {\bibfnamefont {S.}~\bibnamefont {van~den Hoven}},\ and\ \bibinfo {author} {\bibfnamefont {J.~J.}\ \bibnamefont {Renema}},\ }\bibfield  {title} {\bibinfo {title} {Photon-distillation schemes with reduced resource costs based on multiphoton fourier interference},\ }\href@noop {} {\bibfield  {journal} {\bibinfo  {journal} {Physical Review Applied}\ }\textbf {\bibinfo {volume} {23}},\ \bibinfo {pages} {044003} (\bibinfo {year} {2025})}\BibitemShut {NoStop}%
\bibitem [{\citenamefont {Chen}\ \emph {et~al.}(2024)\citenamefont {Chen}, \citenamefont {Peng}, \citenamefont {Guo}, \citenamefont {Gu}, \citenamefont {Ding}, \citenamefont {Liu}, \citenamefont {Zhao}, \citenamefont {You}, \citenamefont {Qin}, \citenamefont {Wang} \emph {et~al.}}]{chen2024heralded}%
  \BibitemOpen
  \bibfield  {author} {\bibinfo {author} {\bibfnamefont {S.}~\bibnamefont {Chen}}, \bibinfo {author} {\bibfnamefont {L.-C.}\ \bibnamefont {Peng}}, \bibinfo {author} {\bibfnamefont {Y.-P.}\ \bibnamefont {Guo}}, \bibinfo {author} {\bibfnamefont {X.-M.}\ \bibnamefont {Gu}}, \bibinfo {author} {\bibfnamefont {X.}~\bibnamefont {Ding}}, \bibinfo {author} {\bibfnamefont {R.-Z.}\ \bibnamefont {Liu}}, \bibinfo {author} {\bibfnamefont {J.-Y.}\ \bibnamefont {Zhao}}, \bibinfo {author} {\bibfnamefont {X.}~\bibnamefont {You}}, \bibinfo {author} {\bibfnamefont {J.}~\bibnamefont {Qin}}, \bibinfo {author} {\bibfnamefont {Y.-F.}\ \bibnamefont {Wang}}, \emph {et~al.},\ }\bibfield  {title} {\bibinfo {title} {Heralded three-photon entanglement from a single-photon source on a photonic chip},\ }\href@noop {} {\bibfield  {journal} {\bibinfo  {journal} {Physical Review Letters}\ }\textbf {\bibinfo {volume} {132}},\ \bibinfo {pages} {130603} (\bibinfo {year} {2024})}\BibitemShut {NoStop}%
\bibitem [{\citenamefont {Cao}\ \emph {et~al.}(2024)\citenamefont {Cao}, \citenamefont {Hansen}, \citenamefont {Giorgino}, \citenamefont {Carosini}, \citenamefont {Zah{\'a}lka}, \citenamefont {Zilk}, \citenamefont {Loredo},\ and\ \citenamefont {Walther}}]{cao2024photonic}%
  \BibitemOpen
  \bibfield  {author} {\bibinfo {author} {\bibfnamefont {H.}~\bibnamefont {Cao}}, \bibinfo {author} {\bibfnamefont {L.}~\bibnamefont {Hansen}}, \bibinfo {author} {\bibfnamefont {F.}~\bibnamefont {Giorgino}}, \bibinfo {author} {\bibfnamefont {L.}~\bibnamefont {Carosini}}, \bibinfo {author} {\bibfnamefont {P.}~\bibnamefont {Zah{\'a}lka}}, \bibinfo {author} {\bibfnamefont {F.}~\bibnamefont {Zilk}}, \bibinfo {author} {\bibfnamefont {J.}~\bibnamefont {Loredo}},\ and\ \bibinfo {author} {\bibfnamefont {P.}~\bibnamefont {Walther}},\ }\bibfield  {title} {\bibinfo {title} {Photonic source of heralded greenberger-horne-zeilinger states},\ }\href@noop {} {\bibfield  {journal} {\bibinfo  {journal} {Physical Review Letters}\ }\textbf {\bibinfo {volume} {132}},\ \bibinfo {pages} {130604} (\bibinfo {year} {2024})}\BibitemShut {NoStop}%
\bibitem [{\citenamefont {Maring}\ \emph {et~al.}(2024)\citenamefont {Maring}, \citenamefont {Fyrillas}, \citenamefont {Pont}, \citenamefont {Ivanov}, \citenamefont {Stepanov}, \citenamefont {Margaria}, \citenamefont {Hease}, \citenamefont {Pishchagin}, \citenamefont {Lema{\^\i}tre}, \citenamefont {Sagnes} \emph {et~al.}}]{maring2024versatile}%
  \BibitemOpen
  \bibfield  {author} {\bibinfo {author} {\bibfnamefont {N.}~\bibnamefont {Maring}}, \bibinfo {author} {\bibfnamefont {A.}~\bibnamefont {Fyrillas}}, \bibinfo {author} {\bibfnamefont {M.}~\bibnamefont {Pont}}, \bibinfo {author} {\bibfnamefont {E.}~\bibnamefont {Ivanov}}, \bibinfo {author} {\bibfnamefont {P.}~\bibnamefont {Stepanov}}, \bibinfo {author} {\bibfnamefont {N.}~\bibnamefont {Margaria}}, \bibinfo {author} {\bibfnamefont {W.}~\bibnamefont {Hease}}, \bibinfo {author} {\bibfnamefont {A.}~\bibnamefont {Pishchagin}}, \bibinfo {author} {\bibfnamefont {A.}~\bibnamefont {Lema{\^\i}tre}}, \bibinfo {author} {\bibfnamefont {I.}~\bibnamefont {Sagnes}}, \emph {et~al.},\ }\bibfield  {title} {\bibinfo {title} {A versatile single-photon-based quantum computing platform},\ }\href@noop {} {\bibfield  {journal} {\bibinfo  {journal} {Nature Photonics}\ }\textbf {\bibinfo {volume} {18}},\ \bibinfo {pages} {603} (\bibinfo {year} {2024})}\BibitemShut {NoStop}%
\bibitem [{\citenamefont {Bayerbach}\ \emph {et~al.}(2023)\citenamefont {Bayerbach}, \citenamefont {D’Aurelio}, \citenamefont {Van~Loock},\ and\ \citenamefont {Barz}}]{bayerbach2023bell}%
  \BibitemOpen
  \bibfield  {author} {\bibinfo {author} {\bibfnamefont {M.~J.}\ \bibnamefont {Bayerbach}}, \bibinfo {author} {\bibfnamefont {S.~E.}\ \bibnamefont {D’Aurelio}}, \bibinfo {author} {\bibfnamefont {P.}~\bibnamefont {Van~Loock}},\ and\ \bibinfo {author} {\bibfnamefont {S.}~\bibnamefont {Barz}},\ }\bibfield  {title} {\bibinfo {title} {Bell-state measurement exceeding 50\% success probability with linear optics},\ }\href@noop {} {\bibfield  {journal} {\bibinfo  {journal} {Science Advances}\ }\textbf {\bibinfo {volume} {9}},\ \bibinfo {pages} {eadf4080} (\bibinfo {year} {2023})}\BibitemShut {NoStop}%
\bibitem [{\citenamefont {Guo}\ \emph {et~al.}(2024)\citenamefont {Guo}, \citenamefont {Zou}, \citenamefont {Ding}, \citenamefont {Zhang}, \citenamefont {Xu}, \citenamefont {Liu}, \citenamefont {Zhao}, \citenamefont {Ge}, \citenamefont {Peng}, \citenamefont {Xu} \emph {et~al.}}]{guo2024boosted}%
  \BibitemOpen
  \bibfield  {author} {\bibinfo {author} {\bibfnamefont {Y.-P.}\ \bibnamefont {Guo}}, \bibinfo {author} {\bibfnamefont {G.-Y.}\ \bibnamefont {Zou}}, \bibinfo {author} {\bibfnamefont {X.}~\bibnamefont {Ding}}, \bibinfo {author} {\bibfnamefont {Q.-H.}\ \bibnamefont {Zhang}}, \bibinfo {author} {\bibfnamefont {M.-C.}\ \bibnamefont {Xu}}, \bibinfo {author} {\bibfnamefont {R.-Z.}\ \bibnamefont {Liu}}, \bibinfo {author} {\bibfnamefont {J.-Y.}\ \bibnamefont {Zhao}}, \bibinfo {author} {\bibfnamefont {Z.-X.}\ \bibnamefont {Ge}}, \bibinfo {author} {\bibfnamefont {L.-C.}\ \bibnamefont {Peng}}, \bibinfo {author} {\bibfnamefont {K.-M.}\ \bibnamefont {Xu}}, \emph {et~al.},\ }\bibfield  {title} {\bibinfo {title} {Boosted fusion gates above the percolation threshold for scalable graph-state generation},\ }\href@noop {} {\bibfield  {journal} {\bibinfo  {journal} {arXiv preprint arXiv:2412.18882}\ } (\bibinfo {year} {2024})}\BibitemShut {NoStop}%
\bibitem [{\citenamefont {Hauser}\ \emph {et~al.}(2025)\citenamefont {Hauser}, \citenamefont {Bayerbach}, \citenamefont {D’Aurelio}, \citenamefont {Weber}, \citenamefont {Santandrea}, \citenamefont {Kumar}, \citenamefont {Dhand},\ and\ \citenamefont {Barz}}]{hauser2025boosted}%
  \BibitemOpen
  \bibfield  {author} {\bibinfo {author} {\bibfnamefont {N.}~\bibnamefont {Hauser}}, \bibinfo {author} {\bibfnamefont {M.~J.}\ \bibnamefont {Bayerbach}}, \bibinfo {author} {\bibfnamefont {S.~E.}\ \bibnamefont {D’Aurelio}}, \bibinfo {author} {\bibfnamefont {R.}~\bibnamefont {Weber}}, \bibinfo {author} {\bibfnamefont {M.}~\bibnamefont {Santandrea}}, \bibinfo {author} {\bibfnamefont {S.~P.}\ \bibnamefont {Kumar}}, \bibinfo {author} {\bibfnamefont {I.}~\bibnamefont {Dhand}},\ and\ \bibinfo {author} {\bibfnamefont {S.}~\bibnamefont {Barz}},\ }\bibfield  {title} {\bibinfo {title} {Boosted bell-state measurements for photonic quantum computation},\ }\href@noop {} {\bibfield  {journal} {\bibinfo  {journal} {npj Quantum Information}\ }\textbf {\bibinfo {volume} {11}},\ \bibinfo {pages} {41} (\bibinfo {year} {2025})}\BibitemShut {NoStop}%
\bibitem [{\citenamefont {Fischer}\ \emph {et~al.}(2016)\citenamefont {Fischer}, \citenamefont {M{\"u}ller}, \citenamefont {Lagoudakis},\ and\ \citenamefont {Vu{\v{c}}kovi{\'c}}}]{fischer2016dynamical}%
  \BibitemOpen
  \bibfield  {author} {\bibinfo {author} {\bibfnamefont {K.~A.}\ \bibnamefont {Fischer}}, \bibinfo {author} {\bibfnamefont {K.}~\bibnamefont {M{\"u}ller}}, \bibinfo {author} {\bibfnamefont {K.~G.}\ \bibnamefont {Lagoudakis}},\ and\ \bibinfo {author} {\bibfnamefont {J.}~\bibnamefont {Vu{\v{c}}kovi{\'c}}},\ }\bibfield  {title} {\bibinfo {title} {Dynamical modeling of pulsed two-photon interference},\ }\href@noop {} {\bibfield  {journal} {\bibinfo  {journal} {New Journal of Physics}\ }\textbf {\bibinfo {volume} {18}},\ \bibinfo {pages} {113053} (\bibinfo {year} {2016})}\BibitemShut {NoStop}%
\bibitem [{Note1()}]{Note1}%
  \BibitemOpen
  \bibinfo {note} {Ref. \cite {ollivier2021hong} reports that the approximation $\protect \mathcal {O}\left ( (g^{(2)}(0))^2 \right ) \approx 0$ holds well for $ g^{(2)}(0))^2 < 0.3$.}\BibitemShut {Stop}%
\bibitem [{\citenamefont {White}\ \emph {et~al.}(2025)\citenamefont {White}, \citenamefont {Polino}, \citenamefont {Ghafari}, \citenamefont {Joch}, \citenamefont {Villegas-Aguilar}, \citenamefont {Shalm}, \citenamefont {Verma}, \citenamefont {Huber},\ and\ \citenamefont {Tischler}}]{white2025robust}%
  \BibitemOpen
  \bibfield  {author} {\bibinfo {author} {\bibfnamefont {S.~J.}\ \bibnamefont {White}}, \bibinfo {author} {\bibfnamefont {E.}~\bibnamefont {Polino}}, \bibinfo {author} {\bibfnamefont {F.}~\bibnamefont {Ghafari}}, \bibinfo {author} {\bibfnamefont {D.~J.}\ \bibnamefont {Joch}}, \bibinfo {author} {\bibfnamefont {L.}~\bibnamefont {Villegas-Aguilar}}, \bibinfo {author} {\bibfnamefont {L.~K.}\ \bibnamefont {Shalm}}, \bibinfo {author} {\bibfnamefont {V.~B.}\ \bibnamefont {Verma}}, \bibinfo {author} {\bibfnamefont {M.}~\bibnamefont {Huber}},\ and\ \bibinfo {author} {\bibfnamefont {N.}~\bibnamefont {Tischler}},\ }\bibfield  {title} {\bibinfo {title} {Robust approach for time-bin-encoded photonic quantum information protocols},\ }\href@noop {} {\bibfield  {journal} {\bibinfo  {journal} {Physical Review Letters}\ }\textbf {\bibinfo {volume} {134}},\ \bibinfo {pages} {180802} (\bibinfo {year} {2025})}\BibitemShut {NoStop}%
\bibitem [{\citenamefont {Valivarthi}\ \emph {et~al.}(2020)\citenamefont {Valivarthi}, \citenamefont {Davis}, \citenamefont {Pe{\~n}a}, \citenamefont {Xie}, \citenamefont {Lauk}, \citenamefont {Narv{\'a}ez}, \citenamefont {Allmaras}, \citenamefont {Beyer}, \citenamefont {Gim}, \citenamefont {Hussein} \emph {et~al.}}]{valivarthi2020teleportation}%
  \BibitemOpen
  \bibfield  {author} {\bibinfo {author} {\bibfnamefont {R.}~\bibnamefont {Valivarthi}}, \bibinfo {author} {\bibfnamefont {S.~I.}\ \bibnamefont {Davis}}, \bibinfo {author} {\bibfnamefont {C.}~\bibnamefont {Pe{\~n}a}}, \bibinfo {author} {\bibfnamefont {S.}~\bibnamefont {Xie}}, \bibinfo {author} {\bibfnamefont {N.}~\bibnamefont {Lauk}}, \bibinfo {author} {\bibfnamefont {L.}~\bibnamefont {Narv{\'a}ez}}, \bibinfo {author} {\bibfnamefont {J.~P.}\ \bibnamefont {Allmaras}}, \bibinfo {author} {\bibfnamefont {A.~D.}\ \bibnamefont {Beyer}}, \bibinfo {author} {\bibfnamefont {Y.}~\bibnamefont {Gim}}, \bibinfo {author} {\bibfnamefont {M.}~\bibnamefont {Hussein}}, \emph {et~al.},\ }\bibfield  {title} {\bibinfo {title} {Teleportation systems toward a quantum internet},\ }\href@noop {} {\bibfield  {journal} {\bibinfo  {journal} {PRX Quantum}\ }\textbf {\bibinfo {volume} {1}},\ \bibinfo {pages} {020317} (\bibinfo {year} {2020})}\BibitemShut {NoStop}%
\bibitem [{\citenamefont {Basso~Basset}\ \emph {et~al.}(2019)\citenamefont {Basso~Basset}, \citenamefont {Rota}, \citenamefont {Schimpf}, \citenamefont {Tedeschi}, \citenamefont {Zeuner}, \citenamefont {Covre~da Silva}, \citenamefont {Reindl}, \citenamefont {Zwiller}, \citenamefont {J{\"o}ns}, \citenamefont {Rastelli} \emph {et~al.}}]{basso2019entanglement}%
  \BibitemOpen
  \bibfield  {author} {\bibinfo {author} {\bibfnamefont {F.}~\bibnamefont {Basso~Basset}}, \bibinfo {author} {\bibfnamefont {M.~B.}\ \bibnamefont {Rota}}, \bibinfo {author} {\bibfnamefont {C.}~\bibnamefont {Schimpf}}, \bibinfo {author} {\bibfnamefont {D.}~\bibnamefont {Tedeschi}}, \bibinfo {author} {\bibfnamefont {K.~D.}\ \bibnamefont {Zeuner}}, \bibinfo {author} {\bibfnamefont {S.~F.}\ \bibnamefont {Covre~da Silva}}, \bibinfo {author} {\bibfnamefont {M.}~\bibnamefont {Reindl}}, \bibinfo {author} {\bibfnamefont {V.}~\bibnamefont {Zwiller}}, \bibinfo {author} {\bibfnamefont {K.~D.}\ \bibnamefont {J{\"o}ns}}, \bibinfo {author} {\bibfnamefont {A.}~\bibnamefont {Rastelli}}, \emph {et~al.},\ }\bibfield  {title} {\bibinfo {title} {Entanglement swapping with photons generated on demand by a quantum dot},\ }\href@noop {} {\bibfield  {journal} {\bibinfo  {journal} {Physical Review Letters}\ }\textbf {\bibinfo {volume} {123}},\ \bibinfo {pages} {160501} (\bibinfo {year} {2019})}\BibitemShut {NoStop}%
\bibitem [{\citenamefont {Eckstein}\ \emph {et~al.}(2011)\citenamefont {Eckstein}, \citenamefont {Christ}, \citenamefont {Mosley},\ and\ \citenamefont {Silberhorn}}]{eckstein2011highly}%
  \BibitemOpen
  \bibfield  {author} {\bibinfo {author} {\bibfnamefont {A.}~\bibnamefont {Eckstein}}, \bibinfo {author} {\bibfnamefont {A.}~\bibnamefont {Christ}}, \bibinfo {author} {\bibfnamefont {P.~J.}\ \bibnamefont {Mosley}},\ and\ \bibinfo {author} {\bibfnamefont {C.}~\bibnamefont {Silberhorn}},\ }\bibfield  {title} {\bibinfo {title} {Highly efficient single-pass source of pulsed single-mode twin beams of light},\ }\href@noop {} {\bibfield  {journal} {\bibinfo  {journal} {Physical Review Letters}\ }\textbf {\bibinfo {volume} {106}},\ \bibinfo {pages} {013603} (\bibinfo {year} {2011})}\BibitemShut {NoStop}%
\bibitem [{\citenamefont {Nielsen}\ and\ \citenamefont {Chuang}(2010)}]{nielsen2010quantum}%
  \BibitemOpen
  \bibfield  {author} {\bibinfo {author} {\bibfnamefont {M.~A.}\ \bibnamefont {Nielsen}}\ and\ \bibinfo {author} {\bibfnamefont {I.~L.}\ \bibnamefont {Chuang}},\ }\href@noop {} {\emph {\bibinfo {title} {Quantum computation and quantum information}}}\ (\bibinfo  {publisher} {Cambridge university press},\ \bibinfo {year} {2010})\BibitemShut {NoStop}%
\bibitem [{Note2()}]{Note2}%
  \BibitemOpen
  \bibinfo {note} {In fact, these small differences result in $P_0 = 1 - \eta - \protect \frac {1}{2}\eta g^{(2)}(0) + \protect \frac {1}{2}\eta ^2 g^{(2)}(0)$ and $P_1 = \eta + \protect \frac {1}{2}\eta g^{(2)}(0) - \eta ^2 g^{(2)}(0)$, but they do not affect the expression for the effective indistinguishability error (Eq. \ref {eq:eps_eff}) nor the expressions for the correction formulas (Eqs. \ref {eq:eps_A} and \ref {eq:eps_B}).}\BibitemShut {Stop}%
\end{thebibliography}%

\end{document}